\documentclass[twocolumn]{autart}    

\usepackage{hyperref}
\usepackage{cite}
\usepackage{graphicx} 
\usepackage{epsfig} 
\usepackage{mathptmx} 
\usepackage{algorithm}
\usepackage{algpseudocode}
\usepackage{times} 
\usepackage{amsmath} 
{
  \theoremstyle{plain}
  
}
\usepackage{amssymb}  
\usepackage{xcolor}
\usepackage{subfigure}
\usepackage{lipsum}
\usepackage{balance}
\usepackage{amsmath}
\usepackage{array}
\usepackage{mathtools}

\newtheorem{theorem}{Theorem}

\newtheorem{lemma}{Lemma}
\newtheorem{problem}{Problem}

\newtheorem{definition}{Definition}
\allowdisplaybreaks

\DeclareMathOperator*{\argmin}{arg\,min}
\DeclareMathAlphabet{\pazocal}{OMS}{zplm}{m}{n}
\newcommand{\norm}[1]{\left\lVert#1\right\rVert}

\newcommand{\cR}{\mathcal{R}}
\newcommand{\cN}{\mathcal{N}}

\newcommand{\J}{\pazocal{J}}

\newcommand{\act}{\mathcal{A}}
\newcommand{\inact}{\mathcal{I}}

\begin{document}

\begin{frontmatter}

\title{First-Order Algorithms for Constrained Nonlinear Dynamic Games}


\author[Minneapolis]{Bolei Di}\ead{dixxx047@umn.edu},    
\author[Minneapolis]{Andrew Lamperski}\ead{alampers@umn.edu},               

\address[Minneapolis]{Electrical Engineering Department, 200 Union St SE, Minneapolis, MN 55455, Minneapolis}  

%
%
%
%
%

\begin{abstract}
  This paper presents algorithms for non-zero sum nonlinear
  constrained dynamic games with full information. Such problems
  emerge when multiple players with action constraints and differing
  objectives interact with the same dynamic system.
  They model a wide range of applications including economics, defense, and energy systems.
  We show how to exploit the temporal structure in
  projected gradient and Douglas-Rachford (DR) splitting
  methods. The resulting algorithms converge locally
  to open-loop Nash equilibria (OLNE) at linear rates.
  Furthermore, we extend stagewise Newton method to find a local
  feedback policy around an OLNE. In the of linear dynamics and polyhedral constraints, we show that this local feedback controller is an approximated feedback Nash equilibrium (FNE).
  Numerical examples are provided.
\end{abstract}

\end{frontmatter}

\section{Introduction}
This paper describes numerical algorithms for
finite-horizon, constrained, discrete-time dynamic games with full
information. In this setup, agents with different objectives but
coupled constraints choose inputs to a dynamic system.
Due to coupling in the constraints, we examine generalized Nash
equilibrium problems (GNEP).
The formulation and all results in this paper automatically apply to
standard Nash equilibrium problems when the action constraints are not coupled.
The dynamic system can be naturally discrete-time or emerge from discretization of a differential game \cite{basar2018a, sethi2019differential, bauso2016game, bressan2011noncooperative}. Dynamic games have many applications including pursuit-evasion
\cite{rusnak2005lady}, active-defense
\cite{prokopov2013linear, garcia2014cooperative}, economics
\cite{el2013dynamic} and the smart grid \cite{zhu2012differential}. Despite a wide array of applications, the computational methods for dynamic games are considerably less developed than the single-agent case of optimal control.

Our previous work \cite{2019arXiv190609097D} extended stagewise Newton algorithm and differential dynamic programming
(DDP), which originated from single-agent optimal control, to unconstrained non-zero-sum dynamic games.
We proved that the methods converge quadratically to an open-loop Nash
equilibrium (OLNE), the resulting closed-loop policies are local
feedback $O(\epsilon^2)$-Nash equilibria (FNEs), and that both
algorithms enjoyed a linear complexity with respect to the horizon.
This paper generalizes \cite{2019arXiv190609097D} to the case of
constrained dynamic games.

Below we review generalized Nash equilibrium problems and their
relation to variational inequalities. Additionally, we describe projected gradient and Douglas-Rachford splitting methods, which can solve variational inequalities (VI) problems.
For a more detailed review of relevant game theory
literature, see \cite{2019arXiv190609097D}.


General Nash equilibrium problems (GNEP) study games with
constraints that couple different players' strategies
\cite{facchinei2007generalized_survey}. For games with continuous
variables, necessary conditions for the solution of a GNEP can be formulated as a
variational inequality (VI).
These inequalities can be solved via generic VI methods or classic feasibility problem methods, such as Newton's method \cite{facchinei2009generalized} or others
\cite{bigi2018nonlinear}. 
First order VI methods require calculation of gradients, while second
order methods require inverting Hessian
matrices. Na\"ive implementations of these calculations respectively
require $O(T^2)$ and $O(T^3)$ steps, where $T$ is the number of
stages.
Our stagewise Newton method from \cite{2019arXiv190609097D} exploits
dynamic structure to compute gradients, Hessians, and inverse Hessians
with $O(T)$ complexity. Below, we describe two VI algorithms in
greater detail.

Projected gradient algorithms alternate between taking gradient
descent steps and projecting onto constraints.
They have
been well studied for optimization \cite{nocedal2006numerical} and variational inequality
(VI) problems \cite{facchinei2007finite}, and proven to converge
linearly in both cases.
This paper describes how the projected gradient method can be applied to
constrained dynamic games, including convergence conditions and convergence rate of the algorithm.

Douglas-Rachford (DR) splitting methods alternate between solving two VI problems when applied to constrained game problems.
Typically one problem corresponds to an unconstrained optimization or game problem, while the other corresponds to projection onto constraints \cite{facchinei2007finite}.
As with the projected gradient method, the DR method converges linearly \cite{giselsson2017tight}.
 In optimization, the DR method is closely related to the alternating direction method of
 multipliers (ADMM) \cite{giselsson2016linear} and has been extended
 to (single-agent) optimal control problems \cite{o2013splitting, stathopoulos2016operator}.

 \subsection{Our Contribution}
We describe efficient implementations of projected gradient and Douglas-Rachford
splitting methods for computing OLNEs of constrained dynamic games.
In both cases, we show how to
exploit temporal structure to achieve iterations of $O(T)$ complexity.
Finally, we adapted the stagewise Newton to solve for a feedback
policy around OLNE trajectories. In the case of games with linear
dynamics and polyhedral constraints, we show that it is an approximate
(generalized) feedback Nash equilibrium.


\subsection{Paper Outline}
The paper can be divided into three major parts.
The first part,
Section~\ref{sec:background}, describes the problem formulations and
solution concepts. The second part, Section~\ref{sec:proj_grad} on
the projected gradient method and Section~\ref{sec:os} on the
Douglas-Rachford splitting method,
describe methods for computing local OLNE solutions.
The third part focuses on local feedback Nash
equilibria. It includes Section~\ref{sec:parametric_games}, which lays the
groundwork, and Section~\ref{sec:local_FNE}, which analyzes local FNE for
constrained dynamic games. Numerical examples are offered in
Section~\ref{sec:example} and future extensions discussed in
Section~\ref{sec:conclusion}.


\section{Problem Formulations and Solution Concepts}
\label{sec:background}
We introduce some standard notation, formulate the constrained dynamic
game problems, and introduce the associated solution concepts.

\subsection{Dynamic and Static Game Formulation}
The main problem of interest is a constrained, deterministic, full-information dynamic game with $N$ players of the form below.
\begin{problem}
\label{prob:dynamic_game}
{\it
\textbf{Constrained nonlinear dynamic game} \\
Each player aims to minimize their own cost
\begin{align}
    \label{eq:cost}
    \min_{u_{n,t:}} \quad J_{n,t} (x, u) = \sum_{k = t}^T c_{n, k}(x_k, u_{:,k})
\end{align}
Subject to dynamic constraints $f_{k}(\cdot)$ and extra constraints $g_{n,k}(\cdot)$
\begin{subequations}
    \label{eq:dynamics}
    \begin{align}
    \text{s.t. } \quad
    & x_{k+1} = f_k(x_k, u_{:,k}) \\
    & g_{k}(x_k, u_{:,k}) \leq 0, \\
    & x_t \textrm{ is fixed.} \\
    & k = t, t+1, \dots, T-1.
    \end{align}
\end{subequations}
}
\end{problem}
Here, $0 \leq t \leq T$ is the starting point for a game.
When  $t = 0$, we call it the \textit{full game}, and $t > 0$, a \textit{subgame $t$}.
Here, the state of the system at time $k$ is denoted by $x_k\in
\mathbb{R}^{n_x}$.
Player $n$'s action at time $k$ is given by $u_{n, k} \in \mathbb{R}^{n_{u_{n}}}$.
The vector of all players' actions at time $k$ is denoted $u_{:, k} = [u_{1,k}^\top, u_{2,k}^\top, \ldots, u_{N,k}^\top]^\top \in \mathbb{R}^{n_u}$.
The cost for player $n$ at time $k$ is $c_{n,k}(x_k,u_{:,k})$.
This encodes the fact that the cost for each player can depend on the actions of all the players. We assume that the costs are twice differentiable with locally
Lipschitz Hessians.

In later analysis, some other notation is helpful.
The actions of player $n$ from time $t$ to $T$ in a vector is denoted $u_{n,t:} = [u_{n,t}^\top, u_{n,t+1}^\top, \ldots,
u_{n,T}^\top]^\top$. The vector of actions other than those of
player $n$ from time $t$ to $T$ is denoted by
$u_{-n,t:} = [u_{1,t:}^\top,\ldots,u_{n-1,t:}^\top,u_{n+1,t:}^\top,\ldots,u_{N,t:}^\top]^\top$.
The vector of states from time $t$ to $T$ is denoted
$x_{t:}=[x_t^\top,x_{t+1}^\top,\ldots,x_T^\top]^\top$ while the vector of all players' actions
from $t$ to $T$ is given by $u_{:,t:} = [u_{1,t:}^\top, u_{2,t:}^\top, \ldots,
u_{N,t:}^\top]^\top$. $x, u$ denote all states and actions collected over all time in a vector.

We denote the set of trajectories $[x_{:,t:}, u_{:,t:}]$ satisfying the dynamic constraint with given $x_t$ as $\pazocal{D}_t(x_t)$ and the extra constraints $\pazocal{G}_t(x_t)$.
Feasible sets of state and actions are denoted $\pazocal{X}_t(x_t)$ and $\pazocal{U}_t(x_t)$ for subgame $t$. The subscript $t$ or initial condition $x_0$ can be suppressed later in this paper indicating values for the full game.

Note that the dynamics are
deterministic, the cost for each player in a subgame can be expressed as a function
of all following actions and given state $x_t$, i.e., $J_{n,t}(x_t,u_{:,t:})$. The dynamics are implicitly substituted to eliminate $x$ when we use such notation.
Problem~\ref{prob:dynamic_game} can be written in an equivalent static game form for ease of notation and analysis.
\begin{problem} \label{prob:static_game}
{\it
\textbf{Static form of game Problem~\ref{prob:dynamic_game}} \\
A static form equivalent to Problem~\ref{prob:dynamic_game} is denoted
\begin{subequations}
    \begin{align}
    \min_{u_{n, t:}} \quad & J_{n,t}(x_t,u_{:,t:}) \\
    \text{s.t. } \quad & u_{:,t:} \in \pazocal{U}_t (x_t)
    \end{align}
\end{subequations}
}
\end{problem}
\subsection{Solution Concept of Games}
We focus on local open-loop Nash equilibria and feedback Nash
equilibria in this paper.
Technically, we are studying generalized Nash equilibria, since players' actions can be coupled in the constraints \cite{facchinei2007generalized}. However, for simplicity,
we will refer to them as Nash equilibria.

\begin{definition}
{\it
\textbf{(Local) open-loop Nash equilibrium} \\
An \emph{open-loop Nash equilibrium (OLNE)} for subgame $t$ of problem
\ref{prob:dynamic_game} and \ref{prob:static_game} with one specific state $x_t$ is a set of actions $u^{\star}_{:,t:} \in \pazocal{U}_t(x_t)$ such that
\begin{equation}
\label{eq:OLNE}
J_{n,t}(x_t, [u_{n,t:}, u_{-n, t:}^{\star}]) \geq J_n(x_t, u^{\star}_{:,t:}), \  n = 1, 2, \ldots, N
\end{equation}
Furthermore, if \eqref{eq:OLNE} only holds for $u_{n,t:} \in \pazocal{U}_{n,t}(x_t)$ in a neighborhood of $u_{n,t:}^\star$, it is called a \emph{local open-loop Nash equilibrium (local-OLNE)}.
}
\end{definition}

Problem~\ref{prob:dynamic_game} and Problem~\ref{prob:static_game} are equivalent in terms of a local OLNE.
An OLNE does not dynamically adjust if the state changes.
In contrast, a feedback Nash equilibrium for dynamic games Problem~\ref{prob:dynamic_game} requires players to be able to measure the state $x_k$ at each step and execute a step-by-step policy $u_{:,k} = \phi^\star_{:,k}(x_k)$ to account for changes in the state. FNE has the valuable property of being subgame perfect \cite{krawczyk2018multistage}.

\begin{definition}
{\it
\textbf{(Local) feedback Nash equilibrium} \\
A collection of feedback policies $u_{n,k} =
\phi_{n,k}^\star(x_k)$, with $\bar u_{n,k} = \phi_{n,k}^*(\bar x_k)$ $\forall n \in \{1, 2, ..., N\}, \forall k \in \{0, 1, \dots, T-1\}$ is said to be a feedback Nash equilibrium of the full game if no player can benefit from changing their policy unilaterally for any subgame $t$, i.e.,
\begin{align}
  \label{eq:fne}
  J_{n,t}(x_t, \phi^\star_{:,t:}) \leq J_{n,t}(x_t, [\phi_{n,t:}, \phi^\star_{-n, t:}]), \ \forall t \in \{0, 1, ..., T\},
\end{align}
where $J_{n,t}(x_t, \phi_{:, t:})$ indicates the total cost of player
$n$ when all players follow policy $\phi_{:,t:}$ for subgame
$t$. All policies should be compatible with the constraints. Furthermone, if \eqref{eq:fne} only holds around $[\bar x, \bar u] \in \pazocal{D}\cap \pazocal{G}$ and
the resulting trajectories remain in a neighborhood of $[\bar x,\bar u]$, it is called a \emph{local feedback Nash equilibrium (local-FNE)}.
}
\end{definition}
This definition only applies to Problem~\ref{prob:dynamic_game} because Problem~\ref{prob:static_game} does not have explicit state information.
In this definition, $[\bar x, \bar u]$ must be an OLNE since
\eqref{eq:fne} could be violated at $\bar x_t$ otherwise. Thus, we focus on local FNEs that are built around OLNEs in this paper.
Ideally, an FNE is solved via the Bellman recursion, which originated from optimal control
problems, and was extended to dynamic games \cite{haurie2012games, krawczyk2018multistage}. Instead of solving the minimizing action at each stage, equilibrium of stagewise games are computed via the following recursion
\begin{subequations}
\label{eq:bellman}
  \begin{align}
  & V_{n,T+1}^\star(x_{T+1}) = 0 \\
  & Q_{n,k}^\star(x_k,u_{:,k}) = c_{n,k}(x_k,u_{:,k}) +
                                  V_{n,k+1}^\star(f_k(x_k,u_{:,k})) \\
  \label{eq:QGame}
  & V_{n,k}^\star(x_k) = \min_{\phi_{n,k}} Q_{n,k}^\star(x_k,\phi_{n,k}(x_k)),\  \text{s.t. } g_{k}(x_k, \phi_{:,k}(x_k)) \leq 0 \\
  & k = 0, 1, ..., T.
  \end{align}
\end{subequations}
Here $V^{\star}_{n, k}(x_k)$ and $Q^{\star}_{n, k}(x_k, u_{:, k})$
are referred to as \textit{equilibrium value functions} for player $n$
at time step $k$. Note that \eqref{eq:QGame} defines a parametric
static game in terms of the $u_{:,k}$ variable at step $k$ and parameterized by $x_k$.
For dynamic games with quadratic costs, linear dynamics and polyhedral constraints, the analytical FNE can be obtained in theory as described in Section~\ref{sec:parametric_games} and Appendix~\ref{app:parametric_games}. In general, this backward recursion is intractable, so we focus on solving it approximately around an OLNE trajectory.
Note that the game ends at $k = T$, and setting
$V_{n,T+1}^\star(x_{T+1}) = 0$ is only for ease of notation and that by construction, $V_{n,t}^*(x_t) =
J_{n,t}(x_t,\phi_{:,t:}^*)$ for $t=0,\ldots,T$.

\subsection{Variational Inequality (VI) Formulation} \label{sec:VI_form}
For continuous-variable games, variational inequalities (VIs) give a
necessary condition for the solution of generalized Nash equilibrium
problems \cite{facchinei2007finite}.
We describe the VI formulation of the full game, which starts at
$t=0$. The formulation for games starting at $t \ge 1$ is simular.
We omit the dependency on $x_0$, stack all of the gradient vector $\frac{\partial J_{n,0}(u, x_0)}{\partial_{u_{n,:}}}$ and define
\begin{align}
\label{eq:JFun}
\pazocal{J}(u) =
\begin{bmatrix}
\frac{\partial J_{1,0}}{\partial u_{1,:}} & \frac{\partial J_{2,0}}{\partial u_{2,:}} & \cdots & \frac{\partial J_{N,0}}{\partial u_{N,:}}
\end{bmatrix}^\top.
\end{align}
Note that $\pazocal{J}(u)$ has the same dimension as $u$. When the
dependence on $x_0$ must be emphasized, we denote the corresponding
function by $\J(x_0,u)$.
The VI
formulation of the full game, Problem~\ref{prob:static_game}, is as following.
\begin{problem}
\label{prob:VI}
{\it
\textbf{VI formulation of static game}
\begin{align}
\text{Find } u^\star \in \pazocal{U} \text{ s.t. } & \pazocal{J}^\top(u) (u - u^\star) \geq 0, \  \forall u \in \pazocal{U}.
\end{align}
}
\end{problem}
Note that $u^\star$ is the solution of VI formulation is only
necessary to $u^\star$ being a local OLNE to
Problem~\ref{prob:static_game}.
To ensure sufficiency,
it should also be checked if each player's action is a local minimizer to their objective.
Two sufficient conditions for player $n$'s cost to be locally
minimized are 1) $\frac{\partial J_n}{\partial u_{n,:}}(u^\star) \neq
0$ and 2) $\frac{\partial J_n}{\partial u_{n,:}}(u^\star) = 0$ and
$J_n(u_{n,:})$ is locally convex.
If $u^\star$ is a local minimizer for all players, then it is also a local OLNE solution to \ref{prob:static_game}.
We refer to this procedure as \textit{playerwise local minimizer check}.

\subsection{Sufficient Conditions for Existence of Solutions}
\label{sec:existence}
A commonly applicable sufficient condition for a solution of the VI problem to exist is that $\pazocal{J}(u)$ is locally strongly monotone and $\pazocal{U}$ is convex.
Another sufficient condition is that, $\pazocal{U}$ is compact convex and $\pazocal{J}(u)$ is continuous on $\pazocal{U}$ \cite{facchinei2007finite}.
In the case if $J_n(x_0, u)$ is convex w.r.t. $u_{n,:}$, the
playerwise local minimizer check described above is satisfied for all
players, and therefore the game has an OLNE solution.
If $\J(x_0,u)$ is continuous in $x_0$ (which is guaranteed by our
differentiability assumption) and locally strongly monotone with
respect to $u$, the dynamic games can be solved for all states near
$x_t$. This guarantees the existence of a local FNE.

\section{The Projected Gradient Method}
\label{sec:proj_grad}

The projected gradient method for monotone VI problems was explained in details in \cite{facchinei2007finite}. We briefly describe the method and its basic application to
Problem~\ref{prob:VI}, which helps find an OLNE for
Problem~\ref{prob:static_game}.
We show how this leads to an algorithm for constrained dynamic games.

\subsection{Projected Gradient Method for VI}
The projection algorithm follows an iterative update
\begin{align}
u^{t+1} = \Pi_{\pazocal{U}} (u^t - \rho \pazocal{J}(u)),
\end{align}
where $\rho$ is a step size and $\Pi$ is the projection operator.
The algorithm converges linearly with constant $(1 + \rho^2 L^2 - 2 \rho \mu)$ when $\pazocal{J}(u)$ is $\mu$-strongly monotone, $L$-Lipschitz, i.e.,
\begin{subequations}
\begin{align}
(\pazocal{J}(u) - \pazocal{J}(v))^\top(u - v) &\geq \mu \norm{u - v}^2 \\
\norm{\pazocal{J}(u) - \pazocal{J}(v)} &\leq L\norm{u - v},
\end{align}
\end{subequations}
and that $\rho L^2 \leq 2\mu$ \cite{facchinei2007finite}. There exists a small $\rho$ that guarantees the linear convergence, although smaller $\rho$ leads to slower convergence. To apply the projection method to constrained dynamic games, we need procedures to compute the gradient $\pazocal{J}(u)$ and the projection onto $\pazocal{U}$.

Given feasible $\bar u$ and $\bar x$ such that $[\bar x, \bar u] \in \pazocal{D}$, the gradient $\pazocal{J}(\bar u)$ can be found efficiently by first performing a backward pass \eqref{eq:grad}\cite{2019arXiv190609097D}, then extracting and stacking corresponding elements in $\frac{\partial J_n(u, x_0)}{\partial u_{:,k}}\Big|_{\bar u}$.
\begin{subequations}
\label{eq:grad}
\begin{align}
\Omega_{n,T+1} &= 0 \\
\Omega_{n,k} &= M_{n,k}^{1x} + \Omega_{n,k+1} A_k \\
\frac{\partial J_n(u, x_0)}{\partial u_{:,k}}\Big|_{\bar u} &= M^{1u}_{n,k} + \Omega_{n,k+1} B_k \\
k &= T, T-1, \dots, 0,
\end{align}
\end{subequations}
where $A_k, B_k, M_k$ are derivatives of the dynamics and cost
evaluated around $\bar x, \bar u$ defined in \eqref{eq:approximations} in Section~\ref{sec:local_FNE}.

The projection onto $\pazocal{U}$ requires solving a constrained optimal control problem with quadratic step costs, which can be solved via classic optimal control methods \cite{rawlings2019model}.
\begin{problem}
\label{prob:con_opticon_u}
{\it
\textbf{Constrained optimal control around nominal trajectory $\bar u$}
}
\begin{subequations}
\begin{align}
    \min_{u} \quad & J(u) = \sum_{k=0}^{T} \norm{u_{:,k} - \bar u_{:,k}}^2 \\
    \text{s.t.} \quad & g_{k}(x_k, u_{:,k}) \leq 0, \\
                    & x_{k+1} = f_k(x_k, u_{:,k})  \\
    & k = 0, 1, ..., T-1.
\end{align}
\end{subequations}
\end{problem}
The notation $J$ is overloaded and is different from that of the formulations of Problem~\ref{prob:dynamic_game} and ~\ref{prob:static_game}.
Note that a trajectory found via projected gradient method is a solution to the VI problem but not necessarily the game. A playerwise local minimizer check needs to be done as discussed in Section~\ref{sec:VI_form}. We summarize the projected gradient method in Algorithm~\ref{alg:pg}.

\begin{algorithm}
\caption{\label{alg:pg} Projected Gradient Method for Constrained Dynamic Games}
\begin{algorithmic}
\State Generate an initial trajectory of actions $\bar u$
\Loop
\State Compute gradient $\pazocal{J}(\bar u)$ according to \eqref{eq:grad}
\State Update $\bar u \leftarrow \bar u - \rho \pazocal{J}(\bar u)$
\State Solve Problem~\ref{prob:con_opticon_u} around $\bar u$, assign the solution to $u$
\State Compare $\bar u$ and $u$ to check convergence
\State Set $\bar u \leftarrow u$
\EndLoop
\State Perform convexity check for $\bar u$
\end{algorithmic}
\end{algorithm}

\section{The Douglas-Rachford Operator Splitting Method}
\label{sec:os}
This section concerns with solving an OLNE with the Douglas-Rachford splitting method, which is an alternative to the projected gradient method. The DR splitting method, in its most general form, finds the vector $w$ that solves a monotone inclusion problem of the form
\begin{align}
  \label{eq:drInclusion}
  0 \in \Theta w + \Phi w
\end{align}
where $\Theta$ and $\Phi$ are two \textit{maximally monotone} operators in this section. The DR algorithm is defined by the iteration
\begin{align}
  w^{t+1} = [(1 - \alpha)I + \alpha R_{\Theta} R_{\Phi}] w^t.
\end{align}
$R_{\Theta} \coloneqq 2 r_{\Theta} - I$ is called the \textit{reflected resolvent} of $\Theta$ where $r_{\Theta}$ is the $resolvent$ of $\Theta$.
The same notation applies for operator $\Phi$. $\alpha$ is a constant such that $\alpha \in (0, 1)$. The key steps are solving for the two operators' resolvents, which we elaborate in the case of constrained dynamic games in this section.
The DR splitting method has been proven to converge linearly in cases when at least one operator has stronger properties such as strong monotonicity and Lipschitz continuity.
See \cite{giselsson2017tight} for more details.

\subsection{VI Reformulation of Static Game with States}
For ease of notation we interpret $[x, u]$ as a column vector vertically stacking $x$ and $u$ in this section. We create an extended gradient $\pazocal{J}_{xu}([x, u])$ prepending $n_x$ of zeros in front of $\pazocal{J}(u)$, i.e.,
\begin{align}
  \pazocal{J}_{xu}([x, u]) =
  \begin{bmatrix}
    \mathbf{0} \\ \pazocal{J}(u)
  \end{bmatrix} \in \mathbb{R}^{n_x + n_u}
\end{align}
and use it to reformulate Problem~\ref{prob:VI} equivalently as
\begin{problem}
  \label{prob:VI_extend}
  {\it
  \textbf{VI formulation of static game with extended gradient} \\
  \begin{subequations}
    \begin{align}
      \text{Find } \quad & [x^\star, u^\star] \in \pazocal{D} \cap \pazocal{G} \\
      \text{ s.t. } \quad & \pazocal{J}_{xu}^\top([x, u]) ([x, u] - [x^\star, u^\star]) \geq 0 \\
      & \forall [x, u] \in \pazocal{D} \cap \pazocal{G}
    \end{align}
  \end{subequations}
  }
\end{problem}

Problem~\ref{prob:VI_extend} is equivalent to an inclusion problem, when $\pazocal{D}$ and $\pazocal{G}$ are convex and $\pazocal{J}_{xu}$ is strongly monotone as explained in Appendix~\ref{app:equivalency}. Note that $x$ is also a decision variable in this formulation.
This VI problem can also be formulated as an inclusion problem such that we can apply the DR-splitting method.

\begin{problem}
  \label{prob:inclusion}
  {\it
    \textbf{Inclusion problem form of the static game Problem~\ref{prob:static_game}}
    \begin{align}
      0 \in (\eta \pazocal{J}_{xu} + \pazocal{N}_{\pazocal{D}} + \pazocal{N}_{\pazocal{G}})([x^\star, u^\star])
    \end{align}
  }
\end{problem}
where $\pazocal{N}_{\pazocal{D}}$ and $\pazocal{N}_{\pazocal{G}}$ indicate the normal cone operators of $\pazocal{N}$ and $\pazocal{G}$, and $\eta$ is a positive regularization constant.
The smaller the $\eta$, the more regularized the problems are, but the slower the convergence would be.
$\eta$ should be tuned so that the relevant problems in Section~\ref{sec:dr_G}, \ref{sec:dr_D} and \ref{sec:dr_J} are either strongly convex of monotone for better solvability.
Note that there are three adding operators in Problem~\ref{prob:inclusion},
in contrast to \eqref{eq:drInclusion} which has only two.
We can single out any one operator and combine the other two to form an inclusion problem with two adding operators and apply the DR algorithm.
The algorithm requires alternately solving two problems that are the resolvents of different operators.
Which combination to chose ultimately depends on which resolvents are easier to solve, which varies with the specific dynamic game and available solvers. Ideally, when both resolvents are analytically solvable, the DR algorithm is preferred.

We summarize the implementation of the DR splitting in Algorithm~\ref{alg:dr} and elaborate the optimization/game problems derived from the resolvents later in this section. We use $y, z$ to indicate a nominal trajectory that follows the same convention as $x, u$.
The notation $J$ is overloaded to indicate costs for these different resolvent problems.
For details regarding how the problems are derived from resolvents, see Appendix~\ref{app:resolvents}.
\begin{algorithm}
  \caption{\label{alg:dr} Douglas-Rachford for Constrained Dynamic Games}
  \begin{algorithmic}
    \State Generate an initial trajectory $\bar x, \bar u$, set $y = \bar x, z = \bar u$
    \State Pick one scheme from Section~\ref{sec:dr_G}, \ref{sec:dr_D} or \ref{sec:dr_J}
    \Loop
    \State Set $y = \bar x, z = \bar u$
    \State Solve the first problem in the selected section with $y, z$, assign the resulting trajectory to $\tilde x, \tilde u$
    \State Reset the intermediate values $\tilde x, \tilde u$ and $y, z$
    \begin{subequations}
      \begin{align}
        [y, z] & \leftarrow 2 [\tilde x, \tilde u] - [y, z]
      \end{align}
    \end{subequations}
    \State Solve the second problem in the selected section with $y, z$, assign the resulting trajectory to $\tilde x, \tilde u$
    \State Reset the intermediate values $\tilde x, \tilde u$ and $y, z$
    \begin{align}
      [y, z] & \leftarrow 2 [\tilde x, \tilde u] - [y, z]
    \end{align}
    \State Compute weighted average
    \begin{align}
      [\bar x, \bar u] \leftarrow (1 - \alpha) {[\bar x, \bar u] + \alpha [y, z]}
    \end{align}
    \State Check for convergence with $\bar x, \bar u$ and $\tilde x, \tilde u$
    \EndLoop
  \end{algorithmic}
\end{algorithm}

\subsection{Singling out $\pazocal{N}_{\pazocal{G}}$} \label{sec:dr_G}
Problem~\ref{prob:inclusion} becomes the following when singling out $\pazocal{N}_{\pazocal{G}}$
\begin{align}
  0 \in ( [\eta \pazocal{J}_{xu} + \pazocal{N}_{\pazocal{D}}] + \pazocal{N}_{\pazocal{G}})([x^\star, u^\star])
\end{align}
The resolvent of $\eta \pazocal{J}_{xu} + \pazocal{N}_{\pazocal{D}}$ corresponds to solving a regularized, unconstrained dynamic game and the resolvent of $\pazocal{N}_{\pazocal{G}}$ is the projection onto $\pazocal{G}$.

\begin{problem}
  \label{prob:reg_dyn_game}
  {\it
  \textbf{Regularized unconstrained dynamic game around nominal trajectory $y, z$}
  }
  \begin{subequations}
    \begin{align}
      \min_{u_{n,:}} \quad & J_n(x, u) = \sum_{k=0}^{T} c_{n, k}(x_k, u_{:,k}) + \frac{1}{2 \eta}\norm{x_{k} - y_{k}}^2 \nonumber  \\
      & \quad \quad \quad \quad \quad \quad \quad \quad \quad \quad \  + \frac{1}{2 \eta}\norm{u_{:,k} - z_{:,k}}^2 \nonumber \\
      & \quad \quad \ \  n = 1, 2, ..., N \\
      \text{s.t.} \quad & x_{k+1} = f_{k}(x_{k}, u_{:, k})
    \end{align}
  \end{subequations}
\end{problem}
This problem can be solved via our previously proposed stagewise Newton or DDP methods with linear complexity in $T$ and quadratic convergence \cite{2019arXiv190609097D}.

\begin{problem}
  \label{prob:proj_G}
  {\it
  \textbf{Projection of a nominal trajectory $y, z$ onto the extra constraints set}
    \begin{align}
      [x, u] = \Pi_{\pazocal{G}}([y, z])
    \end{align}
  }
\end{problem}
When the extra constraints sets are convex at each stage, this projection is equivalent to solving the projection onto a convex set at each stage.

\subsection{Singling out $\pazocal{N}_{\pazocal{D}}$} \label{sec:dr_D}
Problem~\ref{prob:inclusion} becomes the following when singling out $\pazocal{N}_{\pazocal{D}}$
\begin{align}
  0 \in ( [\eta \pazocal{J}_{xu} + \pazocal{N}_{\pazocal{G}}] + \pazocal{N}_{\pazocal{D}})([x^\star, u^\star])
\end{align}
The resolvent of $\eta \pazocal{J}_{xu} + \pazocal{N}_{\pazocal{G}}$ corresponds to solving a series of regularized, constrained static games and the resolvent of $\pazocal{N}_{\pazocal{D}}$ is the projection onto $\pazocal{D}$, which is an optimal control problem.

\begin{problem}
  \label{prob:reg_static_game}
  {\it
  \textbf{Regularized constrained static games around nominal trajectory $y, z$}
  }
  \begin{subequations}
    \begin{align}
      \min_{x_k, u_{:,k}} \quad & J_n(x, u) = \sum_{k=0}^{T} c_{n, k}(x_k, u_{:,k}) + \frac{1}{2 \eta}\norm{x_{k} - y_{k}}^2 \nonumber  \\
      & \quad \quad \quad \quad \quad \quad \quad \quad \quad \quad \  + \frac{1}{2 \eta}\norm{u_{:,k} - z_{:,k}}^2 \\
      \text{s.t.} \quad & g_{k}(x_k, u_{:,k}) \leq 0, \\
      & k = 0, 1, ..., T-1
    \end{align}
  \end{subequations}
\end{problem}
Note that the objectives are not coupled cross time, therefore this is equivalent to solving $T+1$ constrained static games, which can be solved via solving the KKT conditions \cite{facchinei2007generalized} or other static game methods. Note that there is an additional player with decision variable $x_k$ at each stage.

\begin{problem}
  \label{prob:proj_D}
  {\it
  \textbf{Projection of a nominal trajectory $y, z$ onto the dynamics}
  \begin{subequations}
    \begin{align}
      \min_{u} \quad & J(x, u) = \sum_{k=0}^{T} \norm{x_{k} - y_{k}}^2 + \norm{u_{:,k} - z_{:,k}}^2 \\
      \text{s.t.} \quad & x_{k+1} = f_{k}(x_{k}, u_{:, k})
    \end{align}
  \end{subequations}
  }
\end{problem}
This is an optimal control problem with quadratic step cost, which can be solved via classic optimal control methods \cite{Diehl2018}.

\subsection{Singling out $\eta \pazocal{J}_{xu}$} \label{sec:dr_J}
Problem~\ref{prob:inclusion} becomes the following when singling out $\eta \pazocal{J}_{xu}$
\begin{align}
  0 \in ( [\pazocal{N}_{\pazocal{D}} + \pazocal{N}_{\pazocal{G}}] + \eta \pazocal{J}_{xu})([x^\star, u^\star])
\end{align}
The resolvent of $\pazocal{N}_{\pazocal{D}} + \pazocal{N}_{\pazocal{G}}$ corresponds to solving a constrained optimal control problem and the resolvent of $\pazocal{J}_{xu}$ is a series of unconstrained optimization.

\begin{problem}
  \label{prob:con_opticon}
  {\it
  \textbf{Constrained optimal control around nominal trajectory $y, z$}
  }
  \begin{subequations}
    \begin{align}
      \min_{u_{:,k}} \quad & J(x, u) = \sum_{k=0}^{T} \norm{x_{k} - y_{k}}^2 +  \norm{u_{:,k} - z_{:,k}}^2 \\
      \text{s.t.} \quad & g_{k}(x_k, u_{:,k}) \leq 0, \\
                        & x_{k+1} = f_k(x_k, u_{:,k})  \\
      & k = 0, 1, ..., T-1
    \end{align}
  \end{subequations}
\end{problem}
This is an constrained optimal control problem with quadratic step cost, which can be solved via classic optimal control methods \cite{Diehl2018}.

\begin{problem}
  \label{prob:reg_uncon_opt}
  {\it
  \textbf{Regularized unconstrained games around a nominal trajectory $y, z$}
  \begin{subequations}
    \begin{align}
      \min_{x_k, u_{n,k}} \quad & J_n(x, u) = \sum_{k=0}^{T} c_{n, k}(x_k, u_{:,k}) + \frac{1}{2 \eta}\norm{x_{k} - y_{k}}^2 \nonumber  \\
      & \quad \quad \quad \quad \quad \quad \quad \quad \quad \quad \  + \frac{1}{2 \eta}\norm{u_{:,k} - z_{:,k}}^2
    \end{align}
  \end{subequations}
  }
\end{problem}
Because the objectives are not coupled cross time, this becomes $T+1$ unconstrained static games, which can be solved via classic VI methods. Similar to Problem~\ref{prob:reg_static_game}, an additional player with decision variable $x_k$ at each stage is added.

\section{Parametric Games and Feedback Equilibria}
\label{sec:parametric_games}
Solving a parametric game is the backbone of solving an explicit feedback Nash equilibrium of a game. We briefly describe the results of analytically solvable parametric games and dynamic games in this section. Section~\ref{sec:lecqpg} explains the result for linear equality constrained quadratic parametric games, which is directly applied to stagewise Newton method in Section~\ref{sec:local_FNE}.
Section~\ref{sec:lcqdg} describes the form of feedback equilibrium of linearly constrained quadratic dynamic games, which is piecewise affine but can be exponentially complex. We assume each player's cost function $J_n(x, u)$ is continuous and strictly convex throughout this section.

The detailed development of the results can be found in the Appendix~\ref{app:parametric_games}, which is extending the analytical solution of linearly constrained, quadratic objective optimization and optimal control in Chapter 7 of \cite{rawlings2019model}.

\subsection{Linear Equality Constrained Quadratic Parametric Game} \label{sec:lecqpg}
Problem~\ref{prob:lecqpg} is a basic form that is encountered when approximating a constrained dynamic game, where $u = [u_1^\top, u_2^\top, ..., u_N^\top, ]^\top$ collects all players' action and $x$ is a vector parameter.
An FNE policy $u = \phi(x)$ is desired. This game has an explicit analytical solution with simple assumptions.
\begin{problem}
  \it{
    \textbf{Linear equality constrained quadratic parametric game}
    \label{prob:lecqpg}
    \begin{subequations}
      \label{eq:lecqpg}
      \begin{align}
      \min_{u_n} \quad & J_n(x, u) = \frac{1}{2}
      \begin{bmatrix}
        1 \\
        x \\
        u
      \end{bmatrix}^\top
      \begin{bmatrix}
        \Gamma_n^{11} & \Gamma_n^{1x} & \Gamma_n^{1u} \\
        \Gamma_n^{x1} & \Gamma_n^{xx} & \Gamma_n^{xu} \\
        \Gamma_n^{u1} & \Gamma_n^{ux} & \Gamma_n^{uu}
      \end{bmatrix}
      \begin{bmatrix}
        1 \\
        x \\
        u
      \end{bmatrix} \\
      \label{eq:lecqpg_con}
      \text{s.t. } \quad & Wx + Su + p = 0
      \end{align}
    \end{subequations}
  }
\end{problem}
The following lemma describes its solution.

\begin{lemma}
  \label{lem:lecqpg_sol}
  Given playerwise convexity of objective functions, i.e., $\Gamma_{n}^{uu}$ are positive definite,
  Problem~\ref{prob:lecqpg} has a unique affine feedback Nash equilibrium as in \eqref{eq:lecqpg_sol} if $F$ is invertible and $S$ has full row rank.
  \begin{subequations}
    \begin{align}
      \label{eq:lecqpg_sol}
      u^\star =& K x + s \\
      \lambda^\star =& (S F^{-1} S^\top)^{-1}(W - S F^{-1} P) x + \\
      & (S F^{-1} S^\top)^{-1}(p - S F^{-1} H)
    \end{align}
  \end{subequations}
  where
  \begin{subequations}
    \begin{align}
      & K = - F^{-1}S^\top(S F^{-1} S^\top)^{-1}(W - S F^{-1} P) + F^{-1} P \\
      & s = - F^{-1}S^\top(S F^{-1} S^\top)^{-1}(p - S F^{-1} H) + F^{-1} H \\
      & F =
      \begin{bmatrix}
        \Gamma_1^{u_1 u_1} & \Gamma_1^{u_1 u_2} & \hdots & \Gamma_1^{u_1 u_N} \\
        \Gamma_2^{u_2 u_1} & \Gamma_2^{u_2 u_2} & \hdots & \Gamma_2^{u_2 u_N} \\
        \vdots & \vdots & \ddots & \vdots \\
        \Gamma_N^{u_N u_1} & \Gamma_N^{u_N u_2} & \hdots & \Gamma_N^{u_N u_N} \\
      \end{bmatrix} \\
      & P =
      \begin{bmatrix}
        \Gamma_1^{u_1x} \\ \Gamma_2^{u_2x} \\ \vdots \\ \Gamma_N^{u_Nx}
      \end{bmatrix} \quad
      H =
      \begin{bmatrix}
        \Gamma_n^{u_1 1} \\ \Gamma_n^{u_2 1} \\ \vdots \\ \Gamma_n^{u_N 1}
      \end{bmatrix}
    \end{align}
  \end{subequations}
\end{lemma}

\subsection{Linearly Constrained Quadratic Dynamic Games}
\label{sec:lcqdg}
A linearly constrained quadratic dynamic game is formulated as following. It is one of the most complicated form of dynamic games of which we can acquire analytical solution in theory.
\begin{problem}
  \label{prob:lq_dynanic_game}
  {\it
    \textbf{Linearly Constrained Quadratic Dynamic Games}
    \begin{subequations}
      \begin{align}
        \min_{u_n} \quad & J_n(x, u) = \sum_{k=0}^{T} \frac{1}{2}
        \begin{bmatrix}
          1 \\
          x_k \\
          u_{:,k}
        \end{bmatrix}^\top
        \begin{bmatrix}
          \Gamma_{n,k}^{11} & \Gamma_{n,k}^{1x} & \Gamma_{n,k}^{1u} \\
          \Gamma_{n,k}^{x1} & \Gamma_{n,k}^{xx} & \Gamma_{n,k}^{xu} \\
          \Gamma_{n,k}^{u1} & \Gamma_{n,k}^{ux} & \Gamma_{n,k}^{uu}
        \end{bmatrix}
        \begin{bmatrix}
          1 \\
          x_k \\
          u_{:,k}
        \end{bmatrix} \\
        \text{s.t. } \quad & x_{k+1} = A_k x_k + B_k u_{:,k} + b_k \\
        & W_{n,k} x + S_{n,k} u + p_{n,k} \leq 0
      \end{align}
    \end{subequations}
  }
\end{problem}

This problem can be viewed as a static game in $u$ parameterized by
$x$. As developed in detail in Appendix~\ref{app:parametric_games},
the explicit FNE solution found via Bellman recursion to this problem
is a piecewise affine policy, with polyhedral domains, and the value functions are piecewise
quadratic.
However, the required number of polyhedral domains on the space $\pazocal{X}_k$ can
grow exponentially, causing the procedure to be computationally
prohibiting. It is reasonable to believe that the feedback Nash
equilibrium of general constrained dynamic games can be more
complex. This fact drives us to seek local FNE.

\section{Local Feedback Equilibrium}
\label{sec:local_FNE}
OLNEs solved via projected gradient or DR splitting
might not be applicable to systems with noise.
Thus, we aim to find
a local feedback policy around the OLNE that can
accommodate disturbances of the system to a certain degree.
As discussed above,  even for linear-quadratic
dynamic games with polyhedral constraints, explicit feedback strategies can
require exponential
computational complexity.
In this section, we describe a simplified strategy based on linearly
constrained problems.
When the constraints are polyhedral, we prove that the feedback policy is indeed a local
$O(\epsilon^2)-$FNE.

\subsection{Stagewise Newton Method for Local Feedback Policy}
We introduce the stagewise Newton method for computing a local feedback policy for Problem~\ref{prob:dynamic_game}.
The method is based on the stagewise Newton method for unconstrained
dynamic games from our previous work \cite{2019arXiv190609097D}.
The key difference is that, at each step, an approximated linear
equality constrained quadratic game Problem~\ref{prob:lecqpg} is
solved, instead of the unconstrained quadratic game in
\cite{2019arXiv190609097D}.

Suppose a trajectory $\bar x, \bar u$ has been found along with the active constraints at each step.
The set of indices of active constraints in $g_{k}(\bar x_k, \bar u_{:,k})$ is denoted $\bar a_{k}, \forall k$ and we use $g^{\bar a}_{k}(x_k, u_{:,k})$ to indicate the active constraints at $\bar x, \bar u$, therefore $g^{\bar a}_{k}(\bar x_k, \bar u_{:,k}) = 0$.

We inherit the following notation of derivatives for stagewise Newton method from \cite{2019arXiv190609097D}. 
\begin{subequations}
\label{eq:approximations}
\begin{align}
& A_k = \frac{\partial f_k(x_k, u_{:,k})}{\partial x_k} \Big|_{\bar x, \bar u} \quad
    \quad B_{k} = \frac{\partial f_k(x_k, u_{:,k})}{\partial u_{:,k}} \Big|_{\bar x, \bar u} \\
& G_k^l =
    \begin{bmatrix}
    \frac{\partial^2 f^l_k}{\partial x_k^2} & \frac{\partial^2 f^l_k}{\partial x_k   \partial u_{:,k}} \\
    \frac{\partial^2 f^l_k}{\partial u_{:,k} \partial x_k} & \frac{\partial^2 f^l_k}{\partial u_{:,k}^2}
    \end{bmatrix} \Bigg|_{\bar x, \bar{u}}, \ \  l = 1, 2, \ldots, n_x \\
& R_k(\delta x_k, \delta u_{:,k}) =
    \begin{bmatrix}
    \begin{bmatrix}
        \delta x_k \\
        \delta u_{:,k}
    \end{bmatrix}^\top G_k^1
    \begin{bmatrix}
        \delta x_k \\
        \delta u_{:,k}
    \end{bmatrix} \\
    \vdots \\
    \begin{bmatrix}
        \delta x_k \\
        \delta u_{:,k}
    \end{bmatrix}^\top G_k^{n_x}
    \begin{bmatrix}
        \delta x_k \\
        \delta u_{:,k}
    \end{bmatrix}
    \end{bmatrix} \\
& M_{n, k} = \left.
            \begin{bmatrix}
                2 c_{n, k} & \frac{\partial c_{n, k}}{\partial x_k}
                & \frac{\partial c_{n, k}}{\partial u_{:,k}} \\
                \frac{\partial c_{n, k}}{\partial x_k}^\top
                & \frac{\partial^2 c_{n, k}}{\partial x_k^2}
                & \frac{\partial^2 c_{n, k}}{\partial x_k \partial u_{:,k}} \\
                \frac{\partial c_{n, k}}{\partial u_{:,k}}^\top
                & \frac{\partial^2 c_{n, k}}{\partial u_{:,k}\partial x_k}
                & \frac{\partial^2 c_{n, k}}{\partial u_{:,k}^2}
            \end{bmatrix} \right\rvert_{\bar x, \bar u} \\
\nonumber
            & \quad \ \ =
            \begin{bmatrix}
                M_{n, k}^{11} & M_{n, k}^{1x} & M_{n, k}^{1u} \\
                M_{n, k}^{x1} & M_{n, k}^{xx} & M_{n, k}^{xu} \\
                M_{n, k}^{u1} & M_{n, k}^{ux} & M_{n, k}^{uu}
            \end{bmatrix}.
\end{align}
\end{subequations}
We introduce new notation related to the derivatives of the active constraints $a_{k}$, $\forall k \in \{0, 1, 2, ..., T\}$.
\begin{subequations}
\label{eq:approx_active_con_mats}
\begin{align}
W^{\bar a}_{k} &= \frac{\partial g^{\bar a}_{k} (\bar x_k, \bar u_{:,k})}{\partial x} \\
S^{\bar a}_{k} &= \frac{\partial g^{\bar a}_{k} (\bar x_k, \bar u_{:,k})}{\partial u} \\
p^{\bar a}_{k} &= g^{\bar a}_{k} (\bar x_k, \bar u_{:,k})
\end{align}
\end{subequations}

We can form a local linear-quadratic approximation that is still analytically solvable via dynamic programming as following.
\begin{problem} \label{prob:lqApprox_dynamic_game}
{\it
\textbf{Linear-quadratic approximation to Problem~\ref{prob:dynamic_game}}
\begin{subequations}
\label{eq:newton_dp}
\begin{align}
    \label{eq:newton_dp_objective}
    & \min_{u_{n,:}}\frac{1}{2} \sum_{k=0}^T \left(
    \begin{bmatrix}
        1 \\
        \delta x_k \\
        \delta u_{:,k}
    \end{bmatrix}^\top
    M_{n, k}
    \begin{bmatrix}
    1 \\
    \delta x_k \\
    \delta u_{:,k}
    \end{bmatrix}
    + M^{1x}_{n,k} \Delta x_k \right)
\end{align}
subject to
\begin{align}
    & \quad \delta x_0 = 0 \\
    \label{eq:newton_dp_init1}
    & \quad \Delta x_0 = 0 \\
    \label{eq:newton_dp_dynamics0}
    & \quad \delta x_{k+1} = A_k \delta{x}_k + B_k \delta u_{:,k} \\
    \label{eq:newton_dp_dynamics1}
    & \quad \Delta x_{k+1} = A_k \Delta x_k + R_k(\delta x_k, \delta u_{:,k}) \\
  \label{eq:newton_dp_constraints}
    & \quad  W^{\bar a}_{k} \delta x_k + S^{\bar a}_{k} \delta u_{k}
    + p^{\bar a}_{k} = 0 \\
    & \quad k = 0, 1, \ldots, T
\end{align}
\end{subequations}
where the states of the dynamic game are given by $\delta x_k$ and
$\Delta x_k$ as
\begin{subequations}
\label{eq:newton_dp_states}
\begin{align}
    \label{eq:newton_dp_states0}
    & \delta x_k = \sum_{i = 0}^{T} \frac{\partial x_k}{\partial u_{:,i}} \Big|_{\bar x, \bar{u}} \delta u_{:,i} \\
    \label{eq:newton_dp_states1}
    & \Delta x_k^l = \sum_{i = 0}^{T} \sum_{j = 0}^{T} \delta u_{:,i}^\top {\frac{\partial^2 x_k^l}{\partial u_{:,i} \partial u_{:,j}}} \Big|_{\bar x,\bar{u}} \delta u_{:,j}, \  l = 1, 2, \ldots, n_x
\end{align}
\end{subequations}
}
\end{problem}

The following lemma describes the solution to Problem~\ref{prob:lqApprox_dynamic_game}.
\begin{lemma} \label{lem:lqApprox_con_sol}
{\it
The equilibrium value functions found via the Bellman recursion \eqref{eq:bellman} for the dynamic game Problem~\eqref{prob:lqApprox_dynamic_game} are denoted as ${V}^{\bar u}_{n,k}(\cdot)$ and $Q^{\bar u}_{n,k}(\cdot, \cdot)$, which can be expressed as
\begin{subequations}
\label{eq:newton_dp_solution_val_fun}
\begin{align}
\label{eq:newton_dp_solution_state_val_fun}
& V^{\bar u}_{n,k}(\delta x_k, \Delta x_k) = \frac{1}{2} \left(
\begin{bmatrix}
1 \\
\delta x_k
\end{bmatrix}^\top \Lambda_{n,k}
\begin{bmatrix}
1 \\
\delta x_k
\end{bmatrix} + \Omega_{n,k} \Delta x_k \right)
\\
\label{eq:newton_dp_solution_state_action_val_fun}
& Q^{\bar u}_{n,k}(\delta x_k, \Delta x_k, \delta u_{:,k}) = \nonumber \\
& \quad \quad \quad \quad \quad \ \frac{1}{2} \left(
\begin{bmatrix}
1 \\
\delta x_k \\
\delta u_{:,k}
\end{bmatrix}^\top \Gamma_{n,k}
\begin{bmatrix}
1 \\
\delta x_k \\
\delta u_{:,k}
\end{bmatrix} + \Omega_{n,k} \Delta x_k \right)
\end{align}
\end{subequations}
where the matrices $\Lambda_{n,k}$, $\Gamma_{n,k}$, and
$\Omega_{n,k}$ can be computed in a backward pass, which also
finds a local feedback policy of the form $\delta u_{:,k} = K_k \delta
x_k + s_k$ around this trajectory.

The matrices $\Lambda_{n,k}$, $\Gamma_{n,k}$, and $\Omega_{n,k}$ in \eqref{eq:newton_dp_solution_val_fun} are computed recursively by $\Lambda_{n,T+1}=0$, $\Omega_{n,T+1} = 0$, and
\begin{subequations}
\label{eq:newton_dp_matrices}
\begin{align}
\label{eq:newton_dp_solution_omega1}
& \Omega_{n,k} = M_{n,k}^{1x} + \Omega_{n,k+1} A_k \\
\label{eq:newton_dp_solution_D}
& D_{n,k} = \sum_{l=1}^{n_x} \Omega_{n,k+1}^l G_k^l  \\
\label{eq:GammaDef}
& \Gamma_{n,k} = M_{n,k} \nonumber \\
& \ \  +
\begin{bmatrix}
\Lambda_{n, k+1}^{11} & \Lambda_{n,k+1}^{1x}A_k & \Lambda_{n,k+1}^{1x}B_k \\
A_k^\top \Lambda_{n,k+1}^{x1} & A_k^\top \Lambda_{n,k+1}^{xx}A_k + D_k^{xx} & A_k^\top \Lambda_{n,k+1}^{xx} B_k + D_k^{xu} \\
B_k^\top  \Lambda_{n,k+1}^{x1}  & B_k^\top  \Lambda_{n,k+1}^{xx}A_k + D_k^{ux} & B_k^\top \Lambda_{n,k+1}^{xx} B_k + D_k^{uu}
\end{bmatrix} \\
& \quad \ =
\begin{bmatrix}
\Gamma_{n, k}^{11} & \Gamma_{n, k}^{1x} & \Gamma_{n, k}^{1u_1} & \Gamma_{n, k}^{1u_2} & \cdots & \Gamma_{n, k}^{1u_N} \\
\Gamma_{n, k}^{x1} & \Gamma_{n, k}^{xx} & \Gamma_{n, k}^{xu_1} & \Gamma_{n, k}^{xu_2} & \cdots & \Gamma_{n, k}^{xu_N} \\
\Gamma_{n, k}^{u_11} & \Gamma_{n, k}^{u_1x} & \Gamma_{n, k}^{u_1u_1} & \Gamma_{n, k}^{u_1u_2} & \cdots & \Gamma_{n, k}^{u_1u_N} \\
\Gamma_{n, k}^{u_21} & \Gamma_{n, k}^{u_2x} & \Gamma_{n, k}^{u_2u_1} & \Gamma_{n, k}^{u_2u_2} & \cdots & \Gamma_{n, k}^{u_2u_N} \\
\vdots & \vdots & \vdots & \vdots & \ddots & \vdots \\
\Gamma_{n, k}^{u_N1} & \Gamma_{n, k}^{u_Nx} & \Gamma_{n, k}^{u_Nu_1} & \Gamma_{n, k}^{u_Nu_2} & \cdots & \Gamma_{n, k}^{u_Nu_N}
\end{bmatrix} \\
\label{eq:newton_dp_invert}
& F_k =
\begin{bmatrix}
\Gamma_{1k}^{u_1u} \\
\Gamma_{2k}^{u_2u} \\
\vdots \\
\Gamma_{Nk}^{u_Nu}
\end{bmatrix} =
\begin{bmatrix}
\Gamma_{1k}^{u_1u_1} & \Gamma_{1k}^{u_1u_2} & \cdots & \Gamma_{1k}^{u_1u_N} \\
\Gamma_{2k}^{u_2u_1} & \Gamma_{2k}^{u_2u_2} & \cdots & \Gamma_{2k}^{u_2u_N} \\
\vdots & \vdots & \ddots & \vdots \\
\Gamma_{Nk}^{u_Nu_1} & \Gamma_{Nk}^{u_Nu_2} & \cdots & \Gamma_{Nk}^{u_Nu_N}
\end{bmatrix} \\
& P_k =
\begin{bmatrix}
\Gamma_{1k}^{u_1x} \\
\Gamma_{2k}^{u_2x} \\
\vdots \\
\Gamma_{Nk}^{u_Nx}
\end{bmatrix}, \quad
H_k =
\begin{bmatrix}
\Gamma_{1k}^{u_11} \\
\Gamma_{2k}^{u_21} \\
\vdots \\
\Gamma_{Nk}^{u_N1}
\end{bmatrix} \\
\label{eq:newton_strategy}
& K_k = - F_k^{-1}S_k^\top(S_k F_k^{-1} S_k^\top)^{-1}(W_k - S_k F_k^{-1} P_k) + F_k^{-1} P_k \\
& s_k = - F_k^{-1}S_k^\top(S_k F_k^{-1} S_k^\top)^{-1}(p_k - S_k F_k^{-1} H_k) + F_k^{-1} H_k \\
\label{eq:newton_dp_solution_S}
& \Lambda_{n, k} =
\begin{bmatrix}
1 & 0 & s_k^\top \\
0 & I & K_k^\top
\end{bmatrix} \Gamma_{n, k}
\begin{bmatrix}
1 & 0 \\
0 & I \\
s_k & K_k
\end{bmatrix}
\end{align}
\end{subequations}
for $k = T,T-1,\ldots,0$.
}
\end{lemma}
\begin{pf*}{Proof. }
  The stagewise Newton method for unconstrained dynamic game and its solution was proved in \cite{2019arXiv190609097D}. The only difference in the dynamic programming procedure for Problem~\ref{prob:lqApprox_dynamic_game} is that an equality constrained static quadratic game is solved at each time step, resulting in different expressions for the feedback parameters $K_k$ and $s_k$, which are justified by the solution of Problem~\ref{prob:lecqpg}.
  \hfill \qed
\end{pf*}

Note that the differential dynamic programming (DDP) method for unconstrained dynamic games \cite{2019arXiv190609097D} can be adapted to constrained games in a similar fashion, for the sake of completeness of generalizing our previous work. As in the unconstrained dynamic game case, DDP and stagewise Newton do not manifest obvious advantages over each other.

\subsection{Remarks on the Feedback Policy by Stagewise Newton Method}
Stagewise Newton method solves a linear equality constrained
quadratic dynamic game that approximates
Problem~\ref{prob:dynamic_game} around $\bar x, \bar u$ and finds a
local feedback policy. Without constraints $g_{k}(x_k, u_{:,k})$, i.e., setting $S_k = 0, W_k = 0, p_k = 0$, the method reduces to the unconstrained version of stagewise Newton method \cite{2019arXiv190609097D}. However, unlike its counterpart for
unconstrained game, because of the introduction of constraints, the
policy found might not be feasible, therefore we cannot perform the iterative process as in \cite{2019arXiv190609097D}.
We consider in this paper the case when $\bar x,
\bar u$ is an OLNE and the feedback policy found by one backward pass
of the  stagewise Newton
method. In this case, $\bar u$ will be will be a fixed-point
of the Newton iteration and the feedback policy simplifies to $\delta
u_{:,k} = K_k \delta x_k$. Furthermore, the matrices are computed in
$O(T)$ complexity. The possibility of infeasible policy remains and is overcome with tightened constraints as in Section~\ref{sec:poly_con}.

The proposed algorithm neglects the inactive constraints. A problem
arises, in that deviations in the state can cause these neglected
constraints to become violated. A simple method to ensure feasibility
is to tighten the inequality constraints, as is common in model
predictive control \cite{rawlings2019model}. We will see below that in
the case of polyhedral constraints,  the feedback policy is indeed an local $O(\epsilon^2)$-FNE.


\subsection{Problems with Polyhedral Constraints}\label{sec:poly_con}
In this section, we introduce a special class of dynamic problems,
restricting to affine dynamics and polyhedral constraints, such that
the stagewise Newton method can be utilized to find a local
$O(\epsilon^2)$-FNE for a partially tightened problem (defined below). We first introduce a fully tightened version of a polyhedrally constrained linear dynamic game
\begin{problem} \label{prob:tight_lp_dynamic_game}
{\it
\textbf{Game with tightened polyhedral constraints}
\begin{subequations}
    \begin{align}
      \min_{u_{n,:}} \quad & J_{n,t}(x, u) = \sum_{k = t}^T c_{n, k}(x_k, u_{:,k}) \\
    \text{s.t. }
    & x_{k+1} = A_k x_k + B_k u_{:,k} + b_k \\
    & W_k x_k + S_k u_{:,k} + p_k \leq - \gamma_k, \\
                            & x_0 \textrm{ is fixed.}
    \end{align}
\end{subequations}
}
\end{problem}
Here $\gamma_k >0$ are vectors used to tighten the inequality
constraints. As above, say that $\bar x,\bar u$ is an OLNE trajectory. We use superscripts $^{\bar a}$ and $^{\bar i}$ to denote values associated with active and inactive constraints on $\bar x$ and $\bar u$. If we find a local feedback policy
using the stagewise Newton method, it can be shown that it is feasible
for the following partially tightened problem:

\begin{problem} \label{prob:partially_tight_lp_dynamic_game}
{\it
\textbf{Game with partially tightened polyhedral constraints}
\begin{subequations}
  \label{eq:partially_tightened}
    \begin{align}
    \min_{u_{n,:}} \quad & J_{n,t} (x, u) = \sum_{k = t}^T c_{n, k}(x_k, u_{:,k}) \\
    \text{s.t. }
    & x_{k+1} = A_k x_k + B_k u_{:,k} + b_k \\
    & W^{\bar a}_k x_k + S^{\bar a}_k u_{:,k} + p^{\bar a}_k \leq
        - \gamma^{\bar a}_k \\
      \label{eq:inactive_loosened}
& W^{\bar i}_k x_k + S^{\bar i}_k u_{:,k} + p^{\bar i}_k \leq
        0 \\
                            & x_t \textrm{ is fixed.} 
    \end{align}
  \end{subequations}
}
\end{problem}
Recall that the dynamics, constraints, and control law are all
affine. Thus,
when a local variation of the state $\norm{x_k - \bar x} = \epsilon$
happens with a sufficiently small $\epsilon$, the trajectories remain feasible.

The next theorem summarizes this local feedback Nash equilibrium
result. The detailed proof can be found in Appendix~\ref{app:approxEq}.
\begin{theorem} \label{thm:local}{\it
There exists a sufficiently small $\epsilon$, such that if $\norm{x_t
  - \bar x_t} \leq \epsilon$, the OLNE $\bar u, \bar x$, active/inactive
constraints and local feedback policy $\phi_{:,k}(x_k) = K_k x_k +
s_k$ found for Problem~\ref{prob:tight_lp_dynamic_game}, is a local
feedback $O(\epsilon^2)$-Nash equilibrium for
Problem~\ref{prob:partially_tight_lp_dynamic_game}:
\begin{align}
    \label{eq:approx_fne}
    J_{n,t}(x_t, \phi_{:,t:}) \leq J_{n,t}(x_t, [\psi_{n,t:},
  \phi_{-n, t:}]) +O(\epsilon^2), \ \forall t \in \{0, 1, ..., T\}
\end{align}
for any $\psi_{n,:}$ such that the resulting trajectories are
feasible for \eqref{eq:partially_tightened}  and remain in a neighborhood of $[\bar x,\bar u]$.
}
\end{theorem}

\section{Numerical Examples}
\label{sec:example}
We demonstrate the approximated local feedback Nash equilibrium around an OLNE with a common-property fishery resource problem and Douglas-Rachford algorithm with a linear quadratic games with analytically projectable convex constraints.

\subsection{A Common-Property Fishery Resource Problem}
The common-property fishery game was considered in Chapter 13, \cite{sethi2019differential}, which is a classic renewable resource manage problem that dates back to 1970s. The analysis in \cite{sethi2019differential} settled at the conclusion that efficient players will drive some opponents out of the competition and maintain at the bionomic equilibrium with zero sustained economic rent. We demonstrate the dynamic equilibrium of two players jointly utilize the resource for a given period of time.

The game is discretized. Scalars $x_k$ and $u_{1, k}$, $u_{2, k}$ denote the biomass of fish and fishing effort of two players. The fishing effort is constrained by $0 \leq u_{n,k} \leq u_{n}^{\text{max}}$. The dynamics of the system from time $0$ to $T$ is given by
\begin{align}
  x_{k+1} = x_k + \left[ g(x_k) - \sum_{n=1}^{2} q_n u_{n,k} x_k \right] \text{d} t, \ k = 0, 1, ..., T/\text{d}t
\end{align}
where the natural growth rate $g(x_k)$ is
\begin{align}
  g(x_k) = \frac{r}{h^2}(2h x_k - x_k^2) + w_k
\end{align}
$h$ is half the maximal biomass the environment can sustain and $r$ is the maximal growth rate which happens when $x_k = h$. $q_n$ is the catchability coefficient for player $n$. $w_k$ is a I.I.D. Gaussian noise we injected for simulating a noisy system. Step profit of each player is modeled as
\begin{align}
  c_{n,k}(x_k, u_{:,k}) = \left( p_n q_n x_k - e_n \right) u_{n,k} \text{d} t
\end{align}
where $p_n$ is the unit price of landed fish and $e_n$ is the unit cost of effort.

Constants chosen are chosen in favor of player $1$.
\begin{align}
  & u_{1}^{\text{max}} = 0.4, \ u_{2}^{\text{max}} = 0.3, \ r = 8, \ h = 100, \text{d}t = 0.1,\  \nonumber  T = 100, \\
  & q_1 = q_2 = 0.1, \ p_1 = p_2 = 1, \  e_1 = 9, \ e_2 = 11
\end{align}
The \textit{turnpike} for a player is the most profitable level of fish biomass if the player is managing the resource alone. And the \textit{bionomic equilibrium} for a player is the minimal fish biomass that a player can turn a profit.

\begin{figure}[!t]
  \centering
  {
    \includegraphics[width=2.6in]{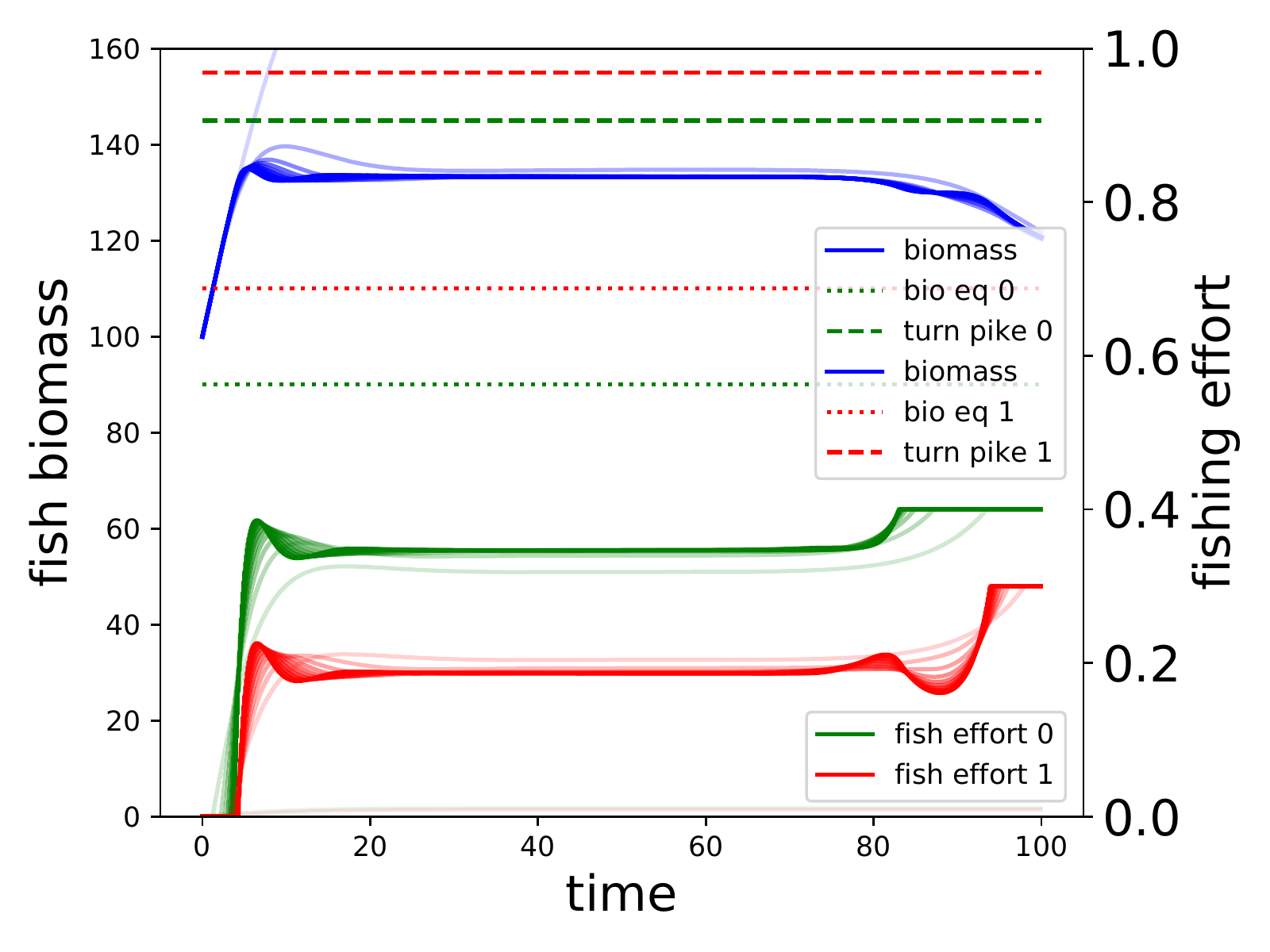}
  }
  \caption{OLNE of projected gradient iterations. 10 trajectories were sampled from 1,000 iterations and shown in Fig. \ref{fig:pg_allinone} with more transparent curves indicating earlier trajectories in the iteration. As can be seen, both players would wait at the beginning for the level of fish biomass to rise even after it passes their bionomic equilibria, since they are managing the resource on a longer term.
  Two players' effort stabilizes in the middle section, which we believe to be the infinite horizon equilibrium for the game, which is not within the scope of this paper.
  In the end, player 1 does not care about longer term profit, so they maximize the effort.
  Player 2 would like to keep the biomass further away their bionomic equilibrium before the final dash, so they reduced effort from time 80 to around 93.}
  \label{fig:pg_allinone}
\end{figure}

\begin{figure}[!t]
\centering
\subfigure[Cumulative Profit]
{
    \includegraphics[width=1.5in]{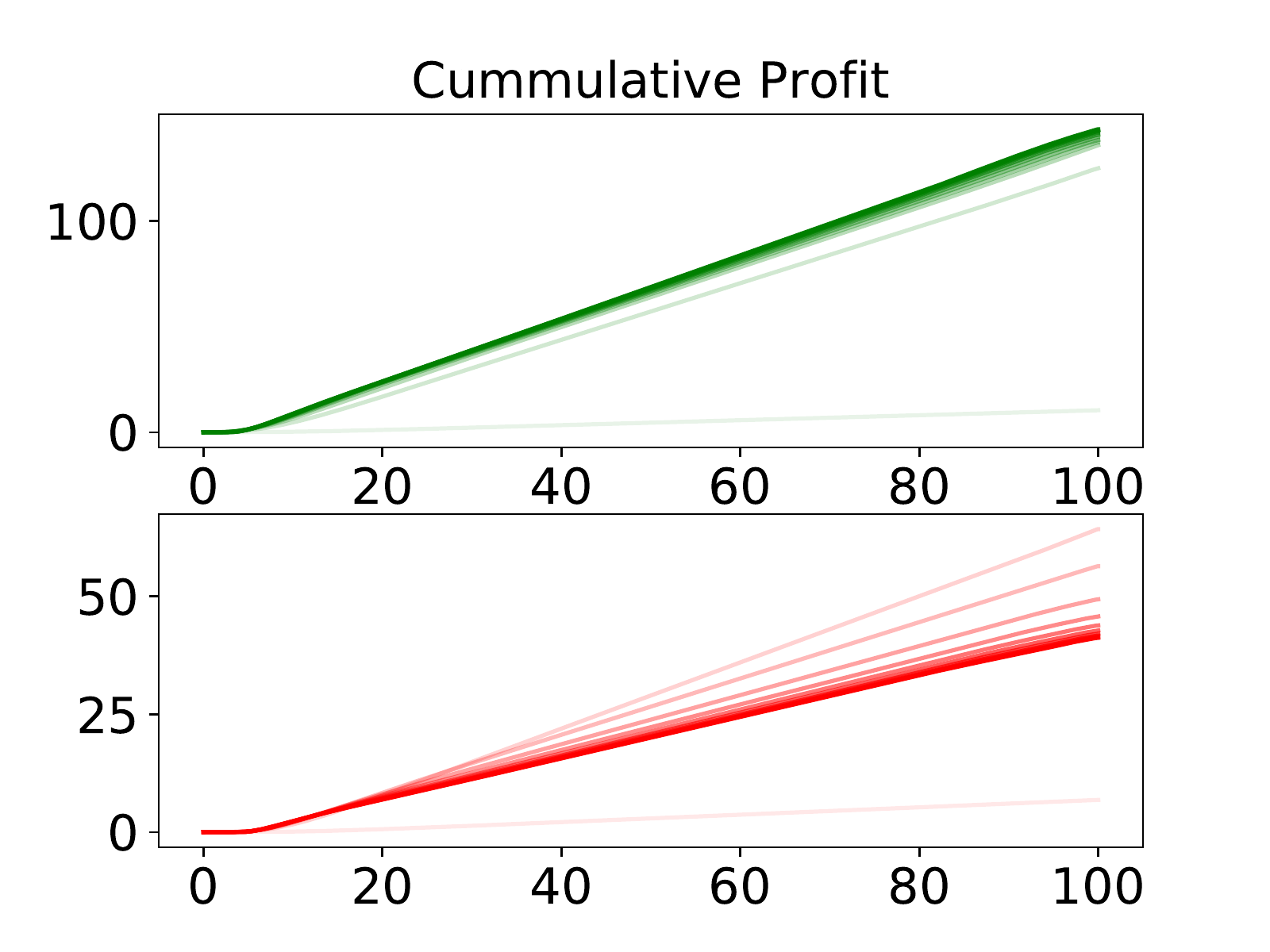}
    \label{fig:pg_CumProf}
}
\subfigure[Convergence]
{
    \includegraphics[width=1.5in]{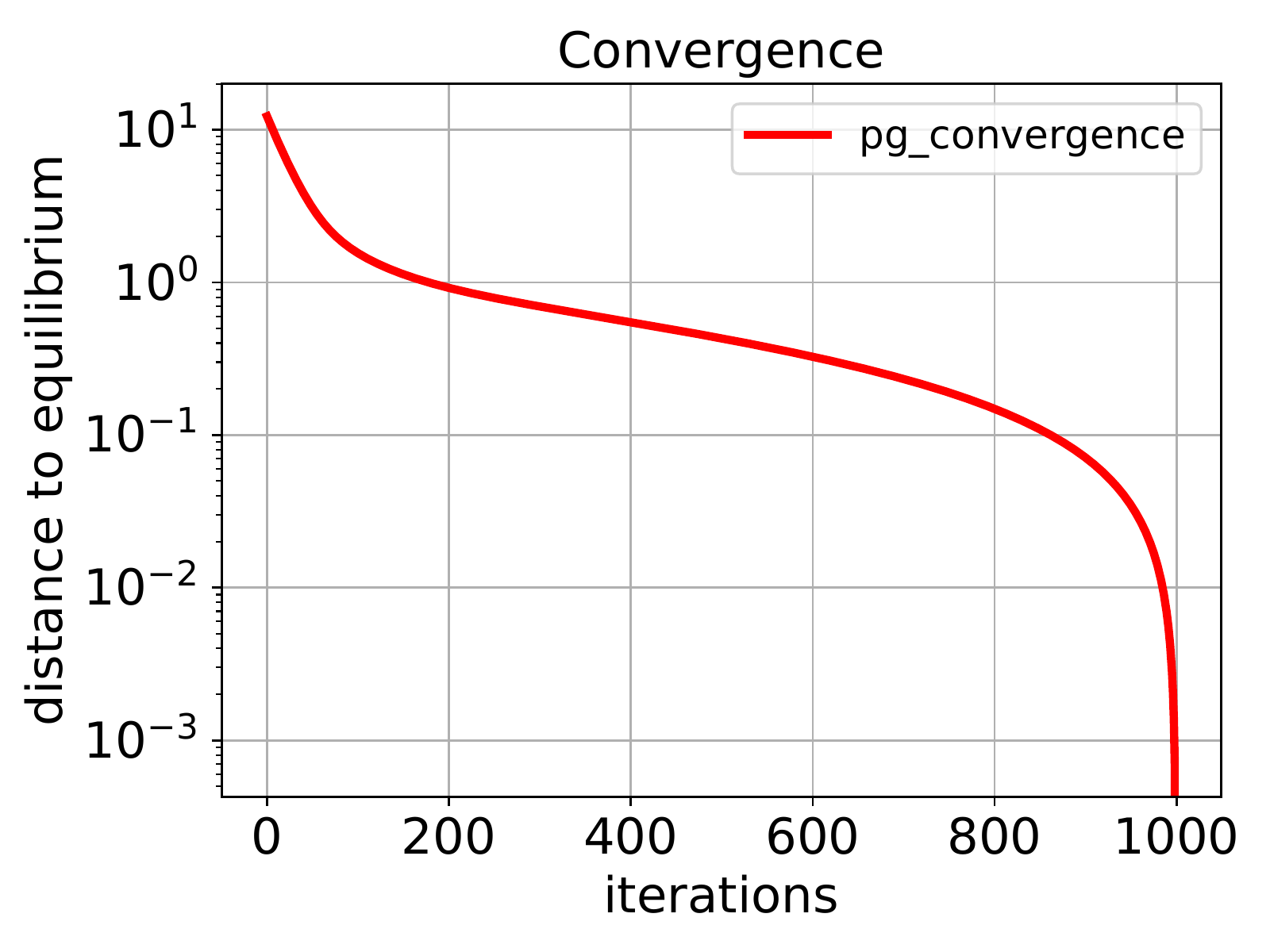}
    \label{fig:pg_convergence.pdf}
}
\caption{Cumulative profit and convergence. The game is formulated favoring player 1, it is not surprising that over iterations, player 1's profit increases while player 2's decreases as in Fig.~\ref{fig:pg_CumProf}. Fig. \ref{fig:pg_convergence.pdf} shows the distance to the final OLNE as the iteration progresses, which fits a typical linear convergence pattern. }
\label{fig:fishery_convergence}
\end{figure}

\begin{figure}[!t]
\centering
\subfigure[OLNE for noisy system]
{
    \includegraphics[width=1.5in]{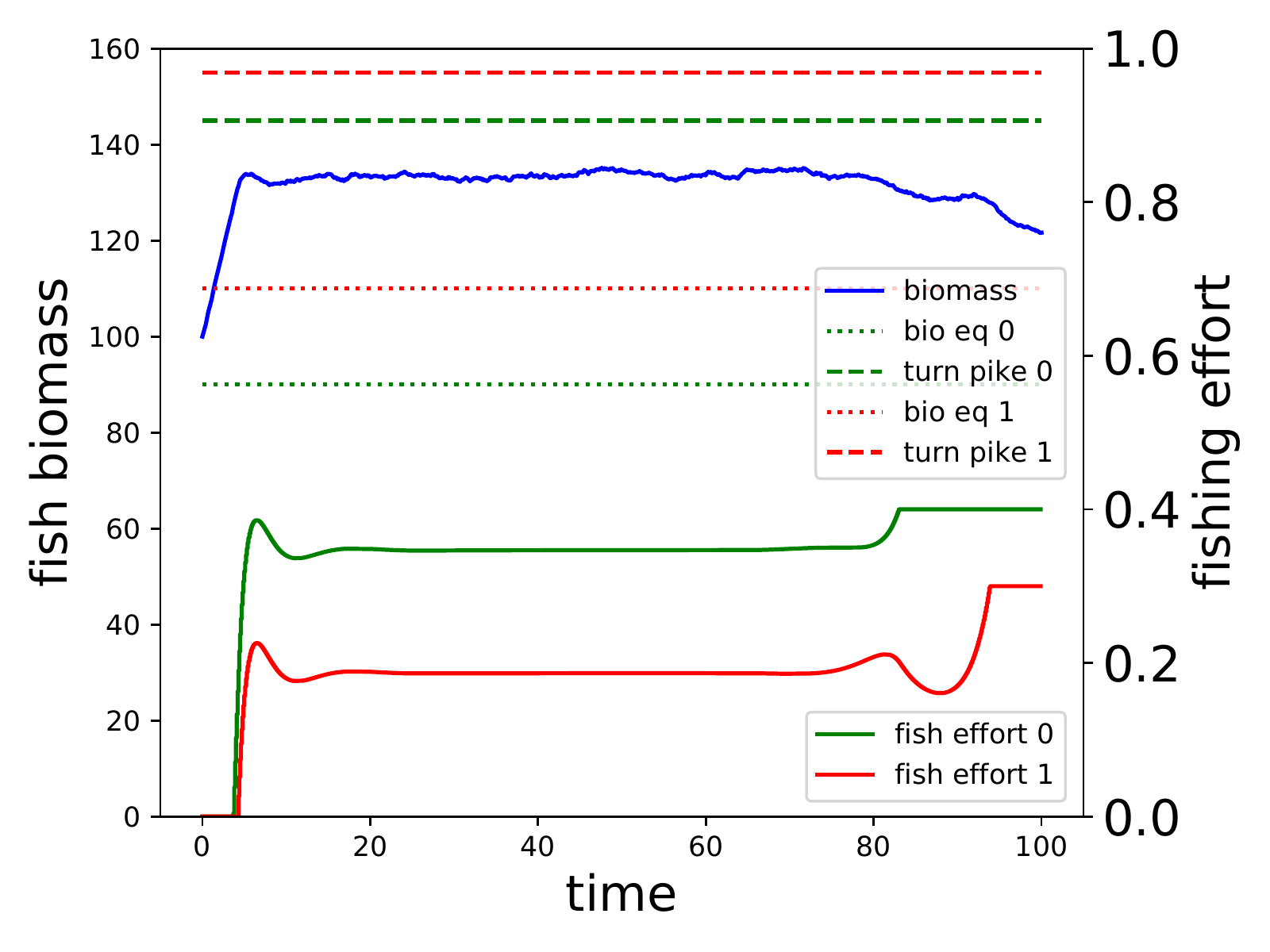}
    \label{fig:pg_noisy_lone}
}
\subfigure[Local FNE for noisy system]
{
    \includegraphics[width=1.5in]{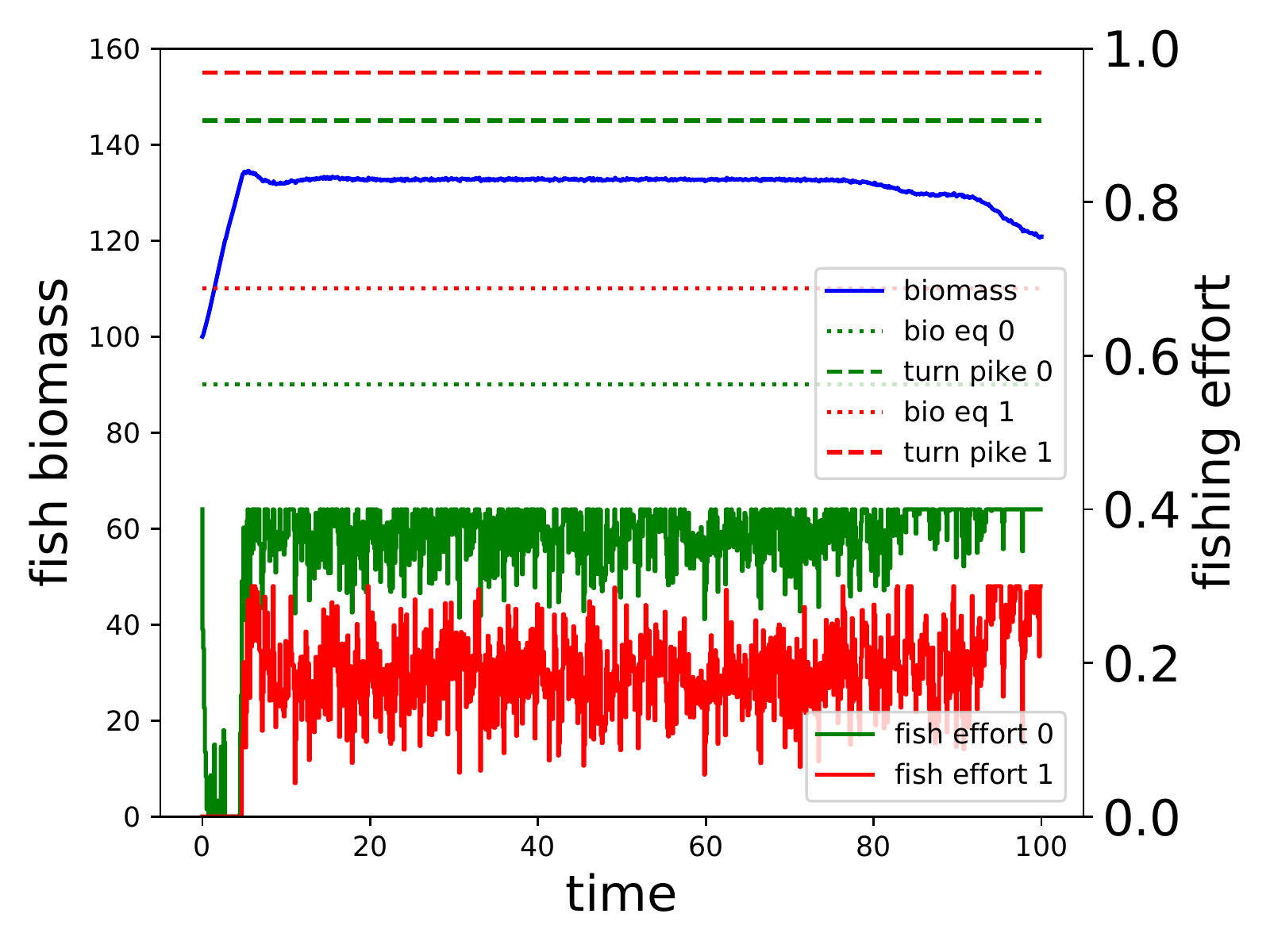}
    \label{fig:pg_noisy_fne}
}
\caption{Local OLNE and FNE for noisy system. Fig. \ref{fig:pg_noisy_lone} shows if the OLNE is blindly applied, the biomass is susceptible to the noise and deviates from the OLNE biomass trajectory. Fig. \ref{fig:pg_noisy_fne} shows the correctional effect of a local FNE with erratic fishing efforts but keeping the biomass smoother and closer to the deterministic OLNE.}
\label{fig:fne_noisy_system}
\end{figure}

We first applied the projected gradient method with stepsize $0.01$ for 1,000 iterations on the deterministic system (neglecting $w_k$), finding the OLNE shown in Fig. \ref{fig:pg_allinone}. The cumulative profit and convergence of actions is summarized in Fig~\ref{fig:fishery_convergence}. We further applied the stagewise Newton method, found the local feedback Nash equilibrium and implemented it for the noisy system, setting the variance of zero-mean Gaussian noise $w_k$ to $\mathbb{E}\{w^2_k\} = 2$. The comparison between OLNE and FNE for noisy system is shown in Fig.~\ref{fig:fne_noisy_system}.

\subsection{Linear Quadratic Game with Convex Constraints}
When a dynamic linear-quadratic(LQ) game has an convex constraint set onto which, the projection is analytically solvable, both problems in Section~\ref{sec:dr_G} can be analytically solved and the Douglas-Rachfor splitting method is preferred.

We demonstrate the DR splitting method on a 2-D locomotion problem with $N=3$ players. Each player directly controls its own location. The system state $x_k \in \mathbb{R}^6$ contains $N$ sets of 2-D coordinates of each player and action $u_{n,k} \in \mathbb{R}^2$ for each players. We use $x_{n,k}$ to denote player $n$'s coordinate at step $k$, naturally we have $x_k = [x_{1, k}, x_{2, k}, x_{3, k}]$, where the variables are all column vectors. The dynamics is simply
\begin{subequations}
  \begin{align}
    x_{k+1} = I_6 x_k + I_6 u_{:,k}, k = 0, 1, ..., T
  \end{align}
\end{subequations}
where $I_6$ is a $6 \times 6$ identity matrix. The initial position $x_0$ and each players target position $x^t_n$ are known. Each player's action is subject to a magnitude constraint $\norm{u_{n,k}} \leq u^{\text{max}}_{n}$. The cost of each player consists of two parts, reaching to the target and conserving its own energy.
\begin{align}
  & c_{n,k} = \norm{x_{n,k} - x^t_n}^2 + 10 \norm{u_{n,k}}^2, \ k = 0, 1, 2,..., T - 1 \\
  & c_{n,T} = 1000\norm{x_{n,T} - x^t_n}^2, \ k = T
\end{align}

It is also required that all three players should meet at $k = 5$, i.e., $x_{1, 5} = x_{2, 5} = x_{3, 5}$, which makes the problem a coupled dynamic game problem rather than 3 separated optimal control problems. The particular constraint also causes the projected gradient method or Algorithm~\ref{alg:pg}, if applied to the problem at hand, to require solving a constrained optimal control problem in each iteration, which needs another iterative procedure. The DR splitting on the other hands, presents two analytically solvable subproblems. In particular, the corresponding Problem~\ref{prob:reg_dyn_game} is solved via stagewise Newton for unconstrained game \cite{2019arXiv190609097D} and Problem~\ref{prob:proj_G} becomes a simple projection onto $\norm{u_{n,k}} \leq u^{\text{max}}_n$.

The parameters are chosen as
\begin{align}
  & T = 10, x_0 = [1, 1, -2, 0, 4, 0], \nonumber \\
  & x^t = [x^t_1, x^t_2, x^t_3] = [4, 12,-2, 10,10, 10], \nonumber \\
  & u^{\text{max}}_{1} = u^{\text{max}}_{2}= u^{\text{max}}_{3} = 2,
\end{align}

\begin{figure}[!t]
  \centering
  \subfigure[Open-loop equilibrium trajectory]
    {
      \includegraphics[width=1.6in]{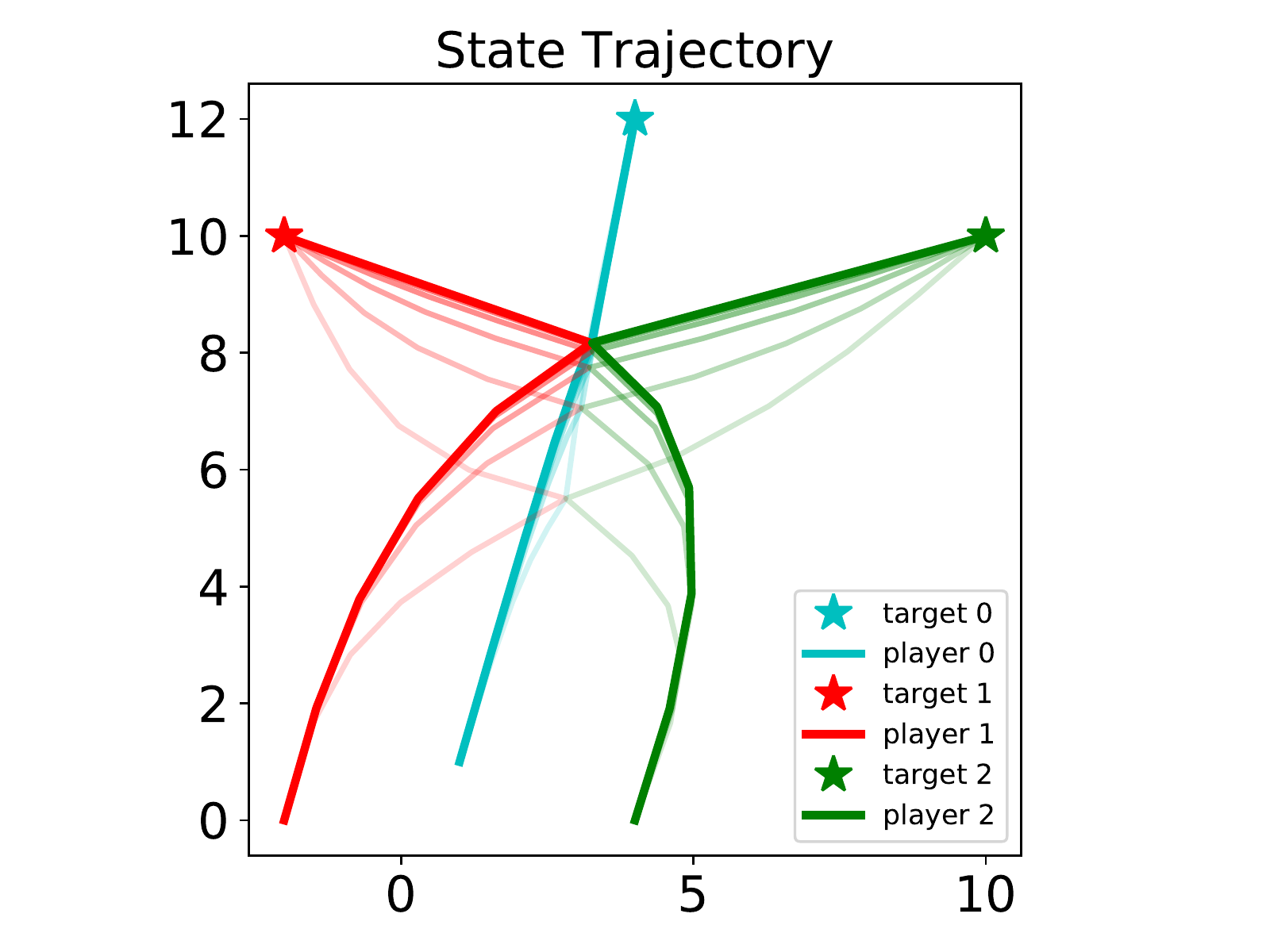}
      \label{fig:trajectory_its}
    }
    \subfigure[Convergence]
    {
      \includegraphics[width=1.5in]{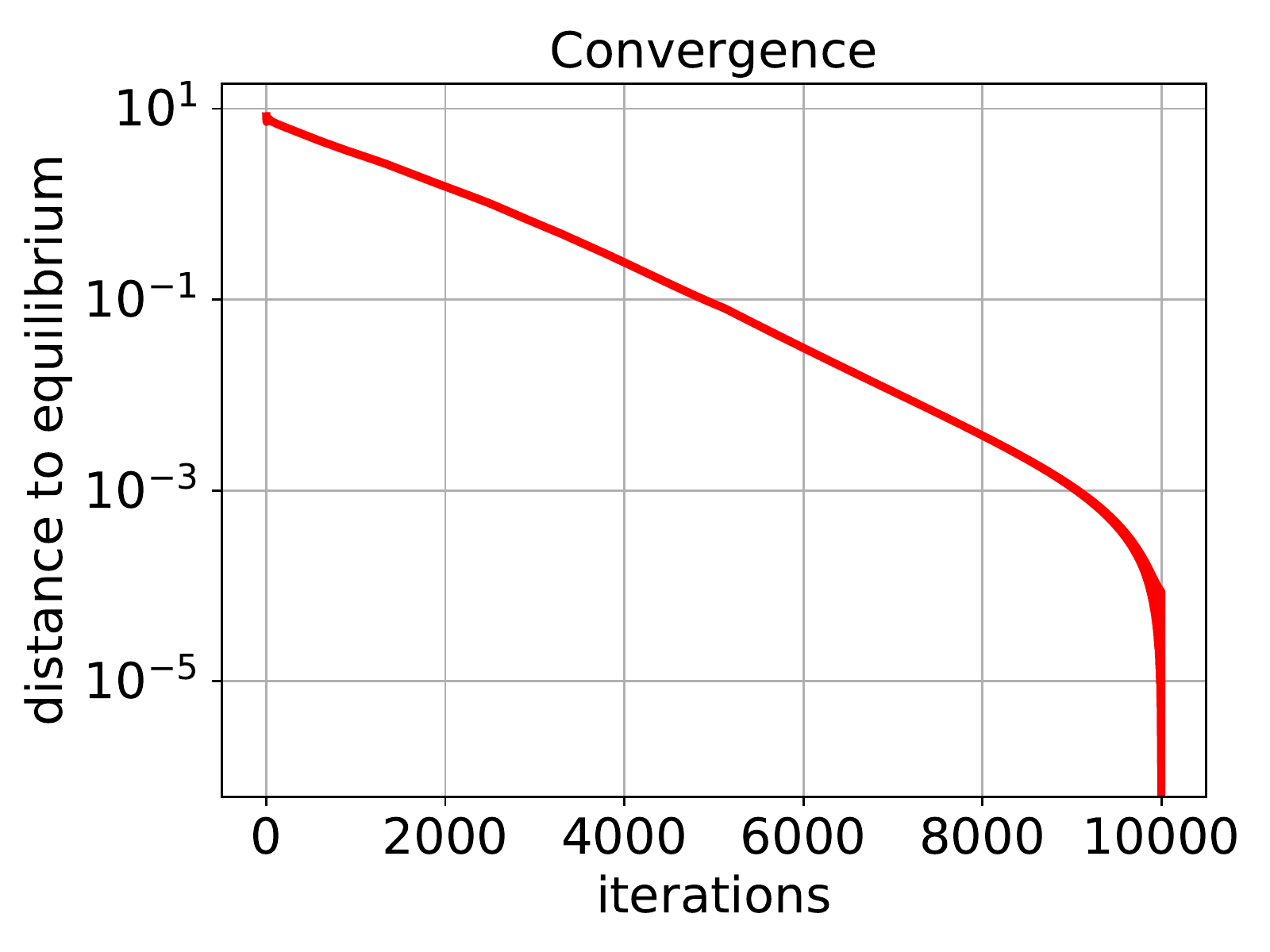}
      \label{fig:convergence.pdf}
    }
  \caption{Douglas-Rachford splitting for dynamic LQ game with convex constraints. 11 trajectories were sampled from 10,000 iterations and shown in Fig. \ref{fig:trajectory_its} with more transparent curves indicating earlier trajectories in the iteration.
  The rendezvous point changed over iteration and all players go straight to target afterwards.
  Fig. \ref{fig:convergence.pdf} shows the distance to the final OLNE as the iteration progresses, which is proof that the algorithm converges. }
  \label{fig:lqConvexCon_convergence}
\end{figure}

We applied the DR splitting method with the splitting scheme in Section~\ref{sec:dr_G}. We found $\eta = 10^{-4}$ and $\alpha = 0.5$ produced stable iteration and solution. $10^4$ iterations were performed and shown in Fig.~\ref{fig:lqConvexCon_convergence}.

\begin{figure}[!t]
  \centering
  \subfigure[Norm of actions]
    {
      \includegraphics[width=1.55in]{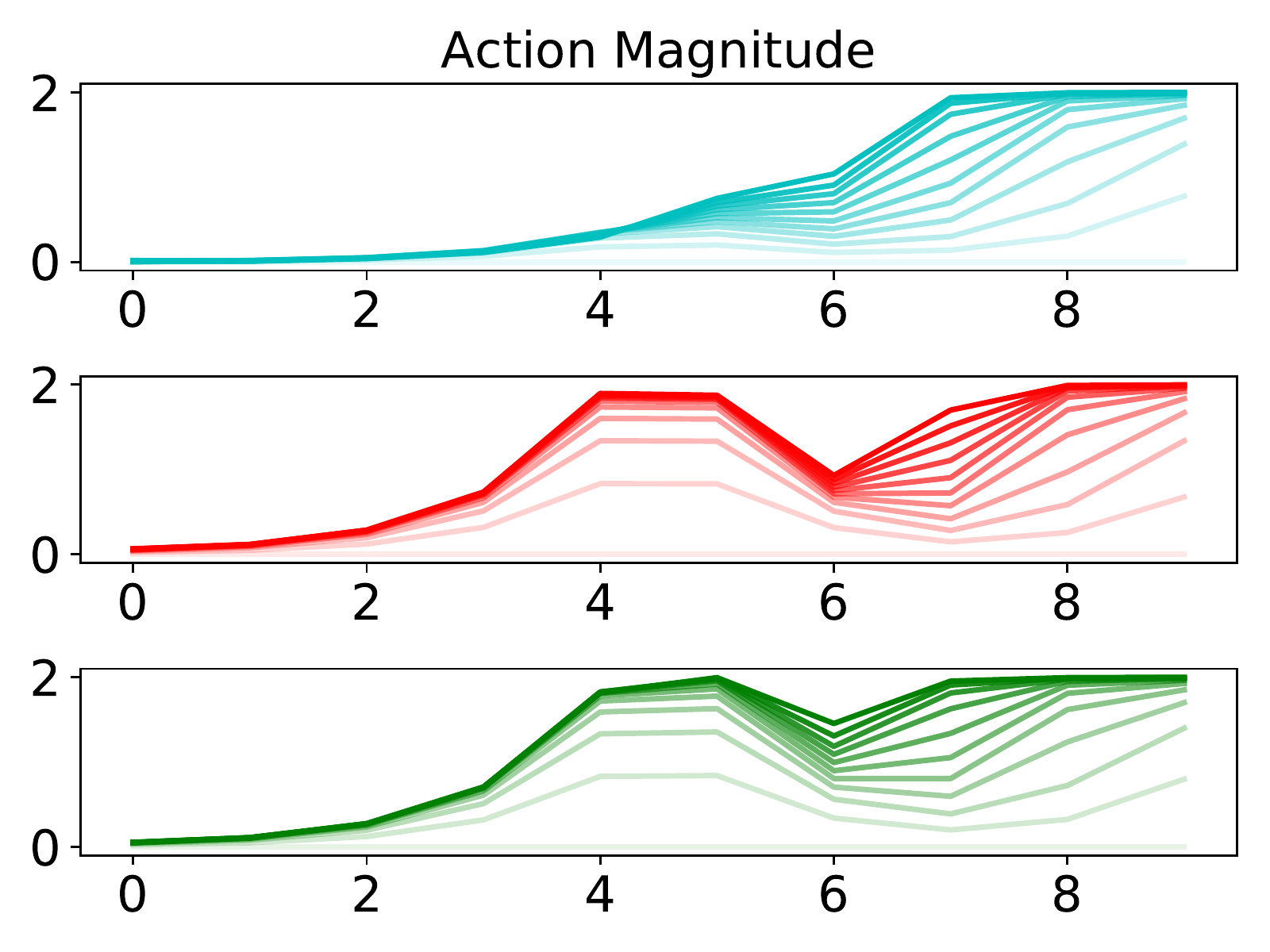}
      \label{fig:action_norm}
    }
    \subfigure[Convergence]
    {
      \includegraphics[width=1.55in]{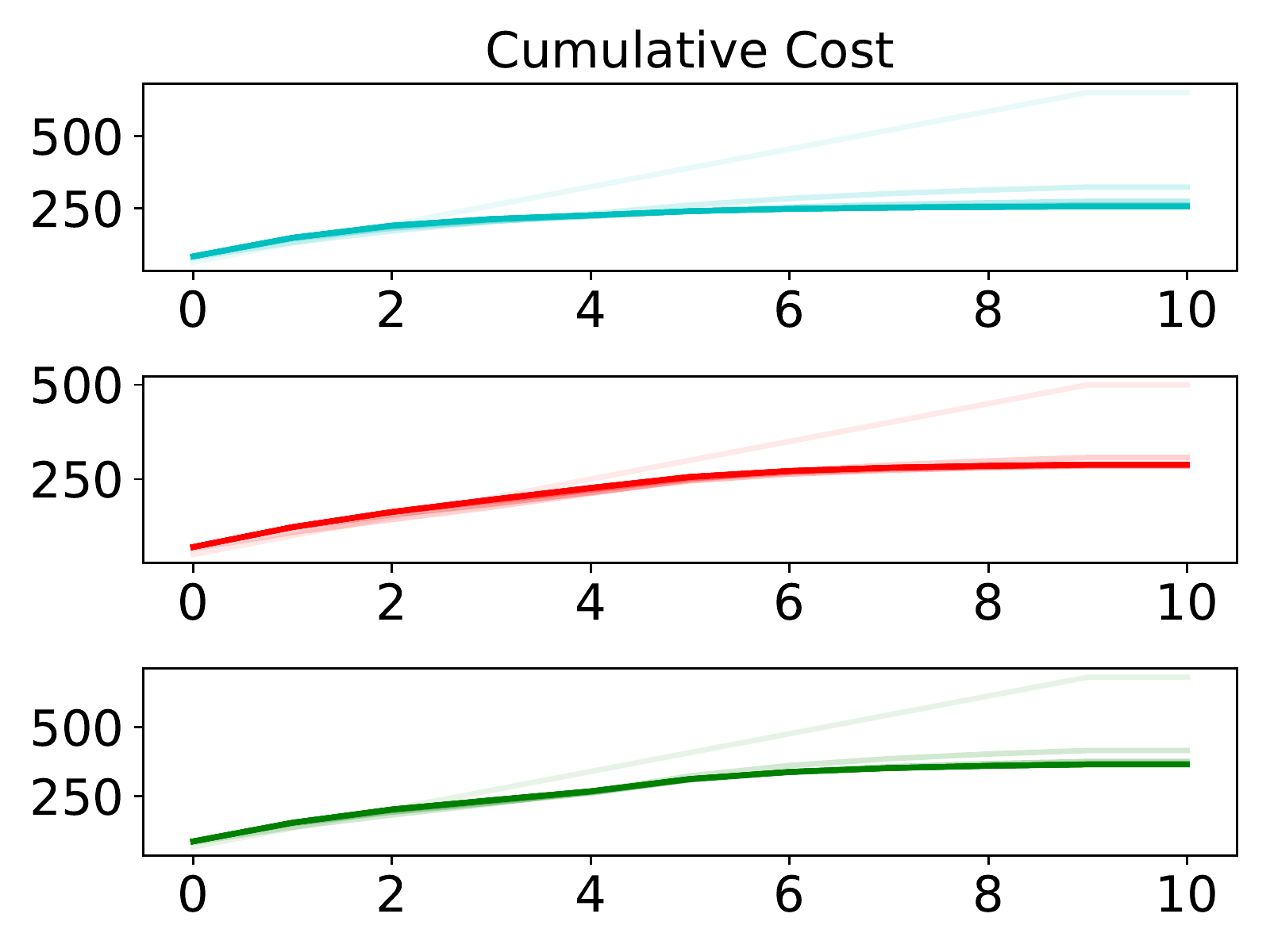}
      \label{fig:cumcost}
    }
  \caption{Magnitudes of action and cumulative cost over iterations. 11 intermediate results were sampled from 10,000 iterations and shown in Fig. \ref{fig:lqConvexCon_action} with more transparent curves indicating earlier trajectories in the iteration.
  The magnitude is bounded by $u^{\text{max}}_n$ as expected in Fig.~\ref{fig:action_norm}.
  Fig. \ref{fig:cumcost} shows the total costs reduce for all players as the rendezvous point changes over iterations.}
  \label{fig:lqConvexCon_action}
\end{figure}

\section{Conclusion and Extensions}
\label{sec:conclusion}
We demonstrated how the projected gradient method and the
Douglas-Rachford algorithm can be used to compute  open-loop Nash
equilibrium of constrained dynamic games. These algorithms converge locally
with a linear rate, and our algorithms require linear iteration
complexity in the horizon. We showed how the OLNE solutions computed
from these methods can be combined with the stagewise Newton method to
find a local feedback strategy. In the case of polyhedrally
constrained games with linear dynamics, was saw that this feedback
policy provides an approximate feedback Nash equilibrium.
The approximation properties of this feedback policy for more general
nonlinearly constrained games is worth further study.
Another promising direction would be to utilize methods from model
predictive control to compute approximate feedback equilibria.
\balance
\bibliographystyle{unsrt}
\bibliography{ref.bib}

\balance

\appendix

\section{Auxiliary Proofs}


\subsection{Problem~\ref{prob:VI_extend} is Equivalent Problem~\ref{prob:inclusion}}
\label{app:equivalency}
Because the normal cone of intersections of convex sets is equal to the sum of normal cones of the convex sets \cite{dontchev2014implicit}, the inclusion problem Problem~\ref{prob:inclusion} is equivalent to
\begin{align}
  \label{eq:incluesion_eq}
  - \eta \pazocal{J}_{xu}([x^\star, u^\star]) \in \pazocal{N}_{\pazocal{D}\cap\pazocal{G}}([x^\star, u^\star])
\end{align}
According to the definition of normal cone
\begin{align}
  & \pazocal{N}_{\pazocal{D}\cap\pazocal{G}}([x^\star, u^\star]) \coloneqq \nonumber \\
  & \quad \{ [y, z] | -[y, z]^\top ([x,y] - [x^\star, u^\star]), \ \forall \ [x, y] \in \pazocal{D}\cap\pazocal{G} \}
\end{align}

\eqref{eq:incluesion_eq} is further equivalent to
\begin{align}
  \eta \pazocal{J}_{xu}([x^\star, u^\star])^\top ([x, u] - [x^\star, u^\star]) \geq 0, \ \forall [x, y] \in \pazocal{D}\cap\pazocal{G}
\end{align}
which is the VI problem Problem~\ref{prob:VI_extend}.

When $\pazocal{J}_{xu}$ is strongly monotone, $\pazocal{D}$ and $\pazocal{G}$ are convex, $\pazocal{J}_{xu}$, $\pazocal{N}_{\pazocal{G}}$ and $\pazocal{N}_{\pazocal{D}}$ are all maximally monotone operators. Because adding operators preserve maximal monotonicity, so we can apply DR splitting algorithm to Problem~\ref{prob:inclusion}.

\subsection{Resolvents of Operators in Problem~\ref{prob:inclusion}}
\label{app:resolvents}
The resolvent of a set-valued/single-valued monotone map $\Phi$ is defined by
\begin{align}
  r_{\Phi} \coloneqq (I + \Phi)^{-1}
\end{align}
which is single-valued and non-expansive \cite{facchinei2007finite}.

\subsubsection{Resolvent of the normal cone of a convex set}
\label{sec:res_normal_cone}
Suppose we have a convex set $\pazocal{X}$. Following the definition, the resolvent of $\pazocal{N}_{\pazocal{X}}(x)$
\begin{subequations}
  \begin{align}
  y = r_{\pazocal{N}_{G}}(x) &= (I + \pazocal{N}_{\pazocal{X}})^{-1}(x) \\
  \end{align}
\end{subequations}
which is equivalent to
\begin{align}
  x - y \in \pazocal{N}_{\pazocal{X}}(y) \\
  (x - y)^\top (z - y) \leq 0, \ \forall z \in \pazocal{X}
\end{align}
This VI problem is equivalent to the optimization of a projection of $x$ onto $\pazocal{X}$
\begin{subequations}
  \begin{align}
    \min_z \quad & y = \norm{x - z} \\
    \text{s.t.} \quad & z \in \pazocal{X}
  \end{align}
\end{subequations}
Therefore the resolvent of a normal cone of a convex set is the projection onto the convex set. This justifies Problem~\ref{prob:proj_G} and \ref{prob:proj_D}.

\subsubsection{Resolvent of a gradient vector}
\label{sec:res_grad}
We study the resolvent of the gradient operator $\eta \pazocal{J}(u)$ of a game as in \eqref{eq:JFun}.
Suppose $u = r_{\eta \pazocal{J}}(x)$, equivalently we have
\begin{subequations}
  \begin{align}
    \label{eq:res_grad_vec}
    u + \eta \pazocal{J}(u) = x
  \end{align}
\end{subequations}
Consider a static game problem
\begin{subequations}
  \begin{align}
    \min_{u_n} \quad & J_n(u) + \frac{1}{2\eta}\norm{u - x}^2
  \end{align}
\end{subequations}
whose Nash equilibrium is equivalent to the solution of
\begin{align}
  \eta \pazocal{J}(u) + u - x = 0
\end{align}
which is equivalent to \eqref{eq:res_grad_vec}. Therefore, the resolvent is equivalent to the solution of a static game.

Based on the arguement of Section~\ref{sec:res_normal_cone} and \ref{sec:res_grad}, the subproblems of Section~\ref{sec:os} can be justified.

\subsection{Generative Cone Condition to Linear Inequalities} \label{app:cone_condition}
\begin{lemma} \label{lem:generative_cone}
  A generative cone condition, where $S$ has full row rank
  \begin{align}
    x \in \text{cone}\{ S^\top \}
  \end{align}
  can be equivalently expressed as
  \begin{align}
    L x \leq 0
  \end{align}
  where
  \begin{align}
    L =
    \begin{bmatrix}
      - (S S^\top)^{-1} S \\
      I - S^\top (S S^\top)^{-1} S \\
      - I + S^\top (S S^\top)^{-1} S
    \end{bmatrix}
  \end{align}
\end{lemma}

\begin{pf*}{Proof. }
  The generative condition is equivalent to
  \begin{subequations}
    \begin{align}
      \label{eq:generative_cone1}
      x &= S^\top \lambda \\
      \lambda &\geq 0
    \end{align}
  \end{subequations}
  which is equivalent to
  \begin{align} \label{eq:generative_cone2}
    \exists M \text{ s.t. } & Mx = \lambda \geq 0 \\
    & (I - S^\top M )x = 0
  \end{align}
  Choose $M = (S S^\top)^{-1} S$, the equivalent conditions can be easily checked
  \begin{subequations}
    \begin{align}
      Mx = (S S^\top)^{-1} S x = (S S^\top)^{-1} S S^\top \lambda = \lambda \geq 0
    \end{align}
  \end{subequations}
  Given the SVD decomposition of $S$
  \begin{align}
    S = U
    \begin{bmatrix}
      \Sigma & 0
    \end{bmatrix}
    \begin{bmatrix}
      V_1^\top \\ V_2^\top
    \end{bmatrix}
  \end{align}
  It is easy to check that
  \begin{align}
    (I - S^\top M ) = I - S^\top (SS^\top)^{-1} S = V_2V_2^\top
  \end{align}
  Since $V_2^\top x = 0$, \eqref{eq:generative_cone2} is true.
  \hfill \qed
\end{pf*}

\subsection{Proof of Theorem~\ref{thm:local}}
\label{app:approxEq}

First we show that the policy produces feasible trajectories.

Recall that $[\bar x,\bar u]$ is an OLNE for the equality-constrained
problem.
Thus, it is a fixed-point of
Newton's method, so that the feedback policy has the form:
\begin{equation}
  \label{eq:localFeedback}
 u_{:,k} = \phi_{:,k}(x_k) =  \bar u_k + K_k (x_k-\bar x_k).
\end{equation}

By construction, \eqref{eq:newton_dp_constraints} ensures that active
constraints for $[\bar x,\bar u]$ remain active for
$[x,u]$.

Now we analyze the inactive constraints. Note that $[\bar x,\bar u]$ satisfies
\begin{equation}
  W^{\bar i}_k \bar x_k + S^{\bar i}_k \bar u_{:,k} + p^{\bar i}_k
  \leq - \gamma_k^{\bar i} <0.
\end{equation}
Furthermore, if $\|x_t - \bar x_t\| \le \epsilon$ the affine dynamics and
polyhedral constraints imply that we must have that
$\|u_{:,k}-\bar u_{:,k}\| = O(\epsilon)$ and $\|x_{k} -\bar x_k \| =
O(\epsilon)$ for all $k \ge t$.
It follows that for sufficiently small
$\epsilon$, the constraint \eqref{eq:inactive_loosened} holds. Thus,
the feedback policy produces a feasible trajectory.

Finally, we show that the approximate feedback equilibrium condition,
\eqref{eq:approx_fne}, holds. Note that left inequality always holds
by construction, so we only need to prove the right inequality.

Fix a player $n$ and a time $t\ge 0$. Assume that the other players are using the
strategy profile $\phi_{-n,t:T}$. Then, with the strategies of the
other players fixed, the policy, $\psi_{n,t:}$, that minimizes $J_{n,t}(x_t,[\psi_{n,t:},\phi_{-n,t:}])$
can be computed from the following optimal control problem:
\begin{subequations}
\label{eq:playerOC}
\begin{align}
  \min_{u_{n,t:T}} \quad & \sum_{k=t}^T
                       c_{n,k}(x_k,(u_{n,k},\phi_{-n,k}(x_k))) \\
   \textrm{s.t.} \quad &
                          u_{-n,k} =\phi_{-n,k}(x_k) \\
    & x_{k+1} = A_k x_k + B_k u_{:,k} + b_k \\
    & W^{\bar a}_k x_k + S^{\bar a}_k u_{:,k} + p^{\bar a}_k \leq
        - \gamma^{\bar a}_k \\
& W^{\bar i}_k x_k + S^{\bar i}_k u_{:,k} + p^{\bar i}_k \leq
        0 \\
                            & x_t \textrm{ is fixed.}
\end{align}
\end{subequations}
The quadraticization around $[\bar x, \bar u]$ of the optimal control
problem \eqref{eq:playerOC} is exactly the same as that of player $n$
in Problem~\ref{prob:partially_tight_lp_dynamic_game}. Thus, when all
other players follow the stagewise Newton strategy, the minimizer $\psi_{n,t:}$ should
be the same as $\phi_{n, t:}$, since player $n$ will have no incentive
to change its strategy on the quadraticized problem. Then, to show
that the strategy is an approximate feedback equilibrium, it suffices
to show that $\phi_{n,t:}$ is approximately optimal. The bound on
optimality follows from the general result on parameterized
optimization from Lemma~\ref{lem:optimizationApproximation} below.
\qed

\subsection{Lemmas on Optimization Approximation}

In this section, we present Lemma~\ref{lem:optimizationApproximation}
which is used to prove Theorem \ref{thm:local}, along with supporting results.

Let $f(x,u)$ be a strongly convex with respect to $u$ in a
neighborhood of $(\bar x,\bar u)$. Assume that $\bar u$ minimizes
$f(\bar x,u)$ with respect to $u$. Let $\hat f(x,u)$ be its quadratic
approximation around the nominal point $(\bar x,\bar u)$. Let
$C(x)=\{u|Wx+Su+p \le 0\}$. Define the policies by

\begin{subequations}
\begin{align}
  \psi(x) &= \argmin_{u\in C(x)} f(x,u) \\
  \rho(x) &= \argmin_{u\in C(x)} \hat f(x,u)
\end{align}
\end{subequations}
Note that $\rho(x)$ has the form
\begin{equation}
  \label{eq:affineLaw}
  u = \rho(x) = K({\act}) x + h({\act}),
\end{equation}
each pair $(K({\act}),h({\act}))$ corresponds to the active indices,
$\act$, of $(x,u)$.

Let $\phi(x) = K(\bar\act)x + h(\bar \act)$, where $\bar \act$ is the
active set of $(\bar u,\bar x)$. Note that in the game context,
this strategy corresponds precisely to the individual players'
approximate strategy computed by stagewise Newton methods.

The eventual goal is to show that
$\phi$ and $\psi$ give similar costs.
As an intermediate result, we
show that $\psi$ and $\rho$ give similar costs.

\begin{lemma}
  \label{lem:sandwich}
    Let $\| x - \bar x \|= \|\delta x\| = \epsilon$. Say that
  $f(x,u)$ has Lipschitz second derivatives and
  is strongly convex with respect to $u$ in a
  neighborhood of $(x,u)$. Further assume that $C(x)$ is non-empty
  for all $x$ in this neighborhood. Then the following bounds hold:
  \begin{multline}
    \label{eq:sandwich}
    f(x,\psi(x)) \le
    f(x,\rho(x))
    \le  f(x,\psi(x)) + O(\epsilon^2)
  \end{multline}
\end{lemma}

\begin{pf*}{Proof. }
  Note that the first inequality of \eqref{eq:sandwich} is immediate
  from the definition of $\psi$. Thus, we focus on the second inequality.

  We will show that $\psi(x)-\rho(x) = O(\epsilon)$. To do this, we
  will show that both functions are  Lipschitz and that $\phi(\bar x)
  = \rho(\bar x) = \bar u$.

  Note that $\psi(x)$ is the
  solution to the following VI
  \begin{equation}
    \nabla_u f(x,u)^\top (v-u) \quad \forall v\in C(x).
  \end{equation}
  By strong convexity and differentiability, $\nabla_u^2 f(x,u)$ is
  positive definite. This implies that for any matrix, $B$ with full
  column rank, we have that $B^\top \nabla_u^2 f(x,u) B$ is also
  positive definite, and thus has positive determinant. Then general
  results on perturbed VIs show that $\psi(x)$ must be Lipschitz. See
  \cite{lu2008variational}. Furthermore, by construction $\psi(\bar x)
  = \bar u$.

  By the same reasoning, we must also have that
  $\rho(x)$ is Lipschitz and again by construction we have that
  $\rho(\bar x) = \bar u$.
  It follows that $\rho(x) = \bar u + O(
  \epsilon)$. Thus
  \begin{align*}
    \rho(x) - \psi(x) = \rho(x) - \bar u + \bar u - \rho(x) = O(\epsilon).
  \end{align*}

  Now we will prove the upper bound.
  For compact notation, let $\tilde u =
  \psi(x)$ and let $u=\rho(x)$.
  Then we have the approximation:
  \begin{equation*}
    f(x,u) = f(x,\tilde u) + \nabla_{u} f(x,\tilde
    u)^\top (u-\tilde u) + O(\epsilon^2).
  \end{equation*}
  By optimality of $\tilde u$ and feasibility of $u$, we have that
  \begin{equation*}
   0 \le \nabla_{u} f(x,\tilde
    u)^\top (u-\tilde u).
  \end{equation*}
  Thus, the proof will be completed if we can
  bound this term above by $O(\epsilon^2)$.

  Let $\J = \nabla_{u} f$.
  Since $\J$ is differentiable and $u-\tilde u = O(\epsilon)$, we have that
  \begin{align*}
    \nabla_u f(x,\tilde u)& = \nabla_u f(x,u) + O(\epsilon) \\
    &= \J(\bar x,\bar u) + \frac{\partial \J}{\partial x}\delta x
      +\frac{\partial \J}{\partial u} \delta u + O(\epsilon),
  \end{align*}
  where the partial derivatives are evaluated at $(\bar x,\bar u)$.

  Thus, the desired bound is given by
   \begin{align*}
    \MoveEqLeft[-1]
    \nabla_{u} f(x,\tilde
    u)^\top (u-\tilde u) \\
 &=
   \left(
\J(\bar x,\bar u) + \frac{\partial \J}{\partial x}\delta x
      +\frac{\partial \J}{\partial u} \delta u
   \right)^\top (u-\tilde u) + O(\epsilon^2) \\
    &\le O(\epsilon^2),
   \end{align*}
   where the second inequality due to optimality of $u$ for the
   corresponding affine VI.
  \qed
\end{pf*}

Now we present the optimization approximation result required for Theorem~\ref{thm:local}.

\begin{lemma}
  \label{lem:optimizationApproximation}
  Assume that $\phi(x) \in C(x)$ and let $\|x-\bar x\| \le
  \epsilon$. Then the following bounds hold.
  \begin{equation}
    f(x,\psi(x)) \le f(x,\phi(x)) \le f(x,\psi(x)) + O(\epsilon^2).
  \end{equation}
\end{lemma}

\begin{pf*}{Proof .}
  The bound on the left holds automatically since $\phi(x)$ is
  feasible and $\psi(x)$ is the corresponding optimal solution.

  Specializing Lemma~\ref{lem:sandwich} to the case of
  optimization implies that $f(x,\psi(x)) =
  f(x,\rho(x))+O(\epsilon^2)$. Thus, it suffices to show that
  \begin{equation}
    f(x,\phi(x)) = f(x,\rho(x)) + O(\epsilon^2).
  \end{equation}

  Let $x(\theta) = (1-\theta)\bar x + \theta x$, let
  $u(\theta) = \rho(x(\theta))$, let $\act(\theta)$ be the
  active set for $(x(\theta),u(\theta))$, and let $\inact(\theta)$
  be the inactive set. We will show that the active sets only switch a
  finite number of times.

  We claim that
  for each subset $\hat \act \subset \{1,\ldots,n_c\}$,
  the set
  $\Theta(\hat \act)=\{\theta | \act(\theta) = \hat \act\}$ is
  convex. In particular, $\Theta(\hat \act)$ must be an interval.

  Proposition 7.10 of \cite{rawlings2019model} implies that the set of
  $x$ such that $(x,\rho(x))$ has active set $\hat \act$ is a
  polyhedron. Since  each $\Theta(\hat \act)$ is the projection of
  the intersection of this set with a line, we must have that
  $\Theta(\hat \act)$ is convex.

  Since there are at most $2^{n_c}$ active sets, it follows that there
  are sets $\act_1,\act_2,\ldots,\act_k$ with $k\le 2^{n_c}$ such that
  $\Theta(\act_i)$ are intervals that partition $[0,1]$. These sets
  can be arranged so that $\theta_i \le \theta_{i+1}$ for $\theta_i
  \in \Theta(\act_i)$ and $\theta_{i+1} \in \Theta(\act_{i+1})$.

  Let $\rho^i$ be the affine strategy corresponding to $\act_i$. Then
  by construction, we have
  \begin{align}
    \label{eq:telescope}
    f(x,\phi(x)) - f(x,\rho(x)) = \sum_{i=1}^{k-1} (f(x,\rho^i(x)) - f(x,\rho^{i+1}(x))).
  \end{align}

  These active sets have the property that either $\act_i \subset \act_{i+1}$
  or $\act_i \supset \act_{i+1}$. Either $\act_i$ must be open on the
  right or $\act_{i+1}$ must be open on the left. Consider the case that
  $\act_i$ is open on the right. (The case of $\act_{i+1}$ open on the
  left is similar.) Then the boundary between $\act_i$
  and $\act_{i+1}$ is a point $\hat \theta \in \act_{i+1}$. By
  continuity, all of the constraints in $\act_i$ must be tight at
  $\hat \theta$. Thus, we must have that $\act_i \subset \act_{i+1}$.



  Consider the
  case that $\act_i \subset \act_{i+1}$. (The other case is similar.)
  In particular, the constraints from $\act_i$ must be active for both
  corresponding VI solutions. Thus, if $S_{\act_i}^+$ is the
  Moore-Penrose pseudoinverse of $S_{\act_i}$ and $R_{\act_i}$ is a matrix with
  full column rank such that $\cR(R_{\act_i})=\cN(S_{\act_i})$, we must have that
  \begin{equation*}
    \rho^{j}(x) = -S_{\act_i}^+(W_{\act_i} x+ p_{\act_i}) +
    R_{\act_i} z_j,
  \end{equation*}
  for $j=i,i+1$, and some vectors $z_j$. It follows that
  \begin{equation}
    \label{eq:policyDiff}
    \rho^{i+1}(x) - \rho^i(x) = R_{\act_i}(z_{i+1}-z_i)
  \end{equation}

  Let $\theta_i\in \Theta(\act_i)$ be some value for which $\act_i$ is
  the active set for $x(\theta_i)=:x^i$, and let $u^i$ be the
  corresponding optimizer. Note that
  $\|\bar x-x(\theta_i)\|\le \epsilon$. It follows that
  \begin{equation*}
    \nabla_u f(x,\rho^i(x)) = \nabla_u \hat f(x^i,u^i) + O(\epsilon).
  \end{equation*}

  Since $u^i$ minimizes $\hat f(x^i,u)$, with $\act_i$ active, we must
  have that
  \begin{equation*}
    \nabla_u \hat f(x^i,u^i)^\top (u^i+R_{\act_i}z-u^i) \ge 0,
  \end{equation*}
  for all sufficiently small $z$. It follows that
  $\nabla_u \hat f(x^i,u^i)^\top R_{\act_i} = 0$.

  Thus, we can get the bound:
  \begin{align}
    \nonumber
    \MoveEqLeft
    f(x,\rho^{i+1}(x)) \\
    \nonumber
 &= f(x,\rho^i(x)) \\
    \nonumber
   & + \nabla_u f(x,\rho^i(x))^\top
                               (\rho^{i+1}(x)-\rho^i(x)) + O(\epsilon^2)
    \\
    \nonumber
 &= f(x,\rho^i(x)) + \\
    \nonumber
    &\nabla_u \hat f(x^i,u^i)^\top R_{\act_i}(z_{i+1}-z_i) +
      O(\epsilon^2) \\
    \label{eq:summandBound}
    &= f(x,\rho^i(x)) + O(\epsilon^2).
  \end{align}

  An analogous argument argument in the case of $\act_i \supset
  \act_{i+1}$ shows that the same bound holds.

  Plugging  \eqref{eq:summandBound} into \eqref{eq:telescope} completes
  the proof.
  \qed
\end{pf*}

\section{Parametric Games and Feedback Equilibrium in Details}
\label{app:parametric_games}
This section contains the detailed development for Section~\ref{sec:parametric_games}. Some problem definitions and lemmas are restated so this section is self-contained for ease of reading.
We study in this section some basic parametric quadratic games with polyhedral constraints to gain insight of the feedback Nash equilibrium of dynamic games in general.
We assume each player's cost function $J_n(x, u)$ is continuous and strictly convex throughout this section.

\subsection{Linear Equality Constrained Quadratic Parametric Game} \label{app:lecqpg}
Problem~\ref{prob:lecqpg_app} is a basic form that is encountered when approximating a constrained dynamic game, where $u = [u_1^\top, u_2^\top, ..., u_N^\top, ]^\top$ collects all players' action and $x$ is a vector parameter.
An FNE policy $u = \phi(x)$ is desired. This game has an explicit analytical solution with simple assumptions.
\begin{problem}
  \it{
    \textbf{Linear equality constrained quadratic parametric game}
    \label{prob:lecqpg_app}
    \begin{subequations}
      \label{eq:lecqpg_app}
      \begin{align}
      \min_{u_n} \quad & J_n(x, u) = \frac{1}{2}
      \begin{bmatrix}
        1 \\
        x \\
        u
      \end{bmatrix}^\top
      \begin{bmatrix}
        \Gamma_n^{11} & \Gamma_n^{1x} & \Gamma_n^{1u} \\
        \Gamma_n^{x1} & \Gamma_n^{xx} & \Gamma_n^{xu} \\
        \Gamma_n^{u1} & \Gamma_n^{ux} & \Gamma_n^{uu}
      \end{bmatrix}
      \begin{bmatrix}
        1 \\
        x \\
        u
      \end{bmatrix} \\
      \label{eq:lecqpg_con_app}
      \text{s.t. } \quad & Wx + Su + p = 0
      \end{align}
    \end{subequations}
  }
\end{problem}
The following lemma describes its solution.

\begin{lemma}
  \label{lem:lecqpg_sol_app}
  Given playerwise convexity of objective functions, i.e., $\Gamma_{n}^{uu}$ are positive definite,
  Problem~\ref{prob:lecqpg_app} has a unique affine feedback Nash equilibrium as in \eqref{eq:lecqpg_sol_app} if $F$ is invertible and $S$ has full row rank.
  \begin{subequations}
    \begin{align}
      \label{eq:lecqpg_sol_app}
      u^\star =& K x + s \\
      \lambda^\star =& (S F^{-1} S^\top)^{-1}(W - S F^{-1} P) x + \\
      & (S F^{-1} S^\top)^{-1}(p - S F^{-1} H)
    \end{align}
  \end{subequations}
  where
  \begin{subequations}
    \begin{align}
      & K = - F^{-1}S^\top(S F^{-1} S^\top)^{-1}(W - S F^{-1} P) + F^{-1} P \\
      & s = - F^{-1}S^\top(S F^{-1} S^\top)^{-1}(p - S F^{-1} H) + F^{-1} H \\
      & F =
      \begin{bmatrix}
        \Gamma_1^{u_1 u_1} & \Gamma_1^{u_1 u_2} & \hdots & \Gamma_1^{u_1 u_N} \\
        \Gamma_2^{u_2 u_1} & \Gamma_2^{u_2 u_2} & \hdots & \Gamma_2^{u_2 u_N} \\
        \vdots & \vdots & \ddots & \vdots \\
        \Gamma_N^{u_N u_1} & \Gamma_N^{u_N u_2} & \hdots & \Gamma_N^{u_N u_N} \\
      \end{bmatrix} \\
      & P =
      \begin{bmatrix}
        \Gamma_1^{u_1x} \\ \Gamma_2^{u_2x} \\ \vdots \\ \Gamma_N^{u_Nx}
      \end{bmatrix} \quad
      H =
      \begin{bmatrix}
        \Gamma_n^{u_1 1} \\ \Gamma_n^{u_2 1} \\ \vdots \\ \Gamma_n^{u_N 1}
      \end{bmatrix}
    \end{align}
  \end{subequations}
\end{lemma}

\begin{pf*}{Proof. }
  The feedback Nash equilibrium can be solved via solving the KKT conditions of all players \cite{facchinei2007generalized}. The Lagrangians can be formulated as
  \begin{subequations}
    \begin{align}
      J_n(x, u) =& \frac{1}{2}
      \begin{bmatrix}
        1 \\
        x \\
        u
      \end{bmatrix}^\top
      \begin{bmatrix}
        \Gamma_n^{11} & \Gamma_n^{1x} & \Gamma_n^{1u} \\
        \Gamma_n^{x1} & \Gamma_n^{xx} & \Gamma_n^{xu} \\
        \Gamma_n^{u1} & \Gamma_n^{ux} & \Gamma_n^{uu}
      \end{bmatrix}
      \begin{bmatrix}
        1 \\
        x \\
        u
      \end{bmatrix} + \\
      & \lambda^\top (Wx + Su + p)
    \end{align}
  \end{subequations}
  Therefore the KKT conditions are
  \begin{subequations}
    \begin{align}
      \label{eq:lagrangianZero}
      \frac{\partial}{u_n} L_n(x, u) &= \Gamma_n^{u_nx} x + \sum_j^{N} \Gamma_n^{u_n u_j} u_j + \Gamma_n^{u_n 1} + (S^{n_c u_n})^\top \lambda \\
      n &= 1, 2, ..., N \\
      \label{eq:constraintHolds}
      0 &= Wx + Su + p
    \end{align}
  \end{subequations}
  Thus, for any $\lambda$, the unique solution \eqref{eq:lagrangianZero} for $u$  is given by $u=-F^{-1}(Px+h+S^\top \lambda)$. Plugging this result into \eqref{eq:constraintHolds} and solving for $\lambda$ gives
  \begin{multline}
    \lambda^\star = (S F^{-1} S^\top)^{-1}(W - S F^{-1} P) x + \\
     (S F^{-1} S^\top)^{-1}(p - S F^{-1} H)
  \end{multline}
  Plugging this back into the expression for $u$ gives the result.
  \hfill \qed
\end{pf*}

\subsection{Linearly Constrained Quadratic Parametric Game}
\label{app:lcqpg}
The more generalized problem with inequality constraints are studied
in this section. Related problems are studied and the piecewise affine
solution was recognized in the variational inequality literature
\cite{lu2008variational, robinson1992normal}. Our analysis focuses on
games which in addition, recognizes the piecewise quadratic value
functions.
We inherit the notation of $u, x, \Gamma, F, P, H$ and $\phi$ from Section~\ref{app:lecqpg}. Such problems serve as a backbone for analyzing FNE for dynamic games when we solve the static game \eqref{eq:QGame} formed by the state-action value function at a stage.
\begin{problem} \label{prob:lcqpg}
  {\it
    \textbf{Linearly constrained quadratic parametric game}
    \begin{subequations}
      \begin{align}
      \min_{u_n} \quad & J_n(x, u) = \frac{1}{2}
      \begin{bmatrix}
        1 \\
        x \\
        u
      \end{bmatrix}^\top
      \begin{bmatrix}
        \Gamma_n^{11} & \Gamma_n^{1x} & \Gamma_n^{1u} \\
        \Gamma_n^{x1} & \Gamma_n^{xx} & \Gamma_n^{xu} \\
        \Gamma_n^{u1} & \Gamma_n^{ux} & \Gamma_n^{uu}
      \end{bmatrix}
      \begin{bmatrix}
        1 \\
        x \\
        u
      \end{bmatrix} \\
      \text{s.t. } \quad & Wx + Su + p \leq 0
      \end{align}
    \end{subequations}
    We use $\pazocal{X}$ and $\pazocal{U}$ to indicate the feasible sets of $x$ and $u$, i.e.,
    \begin{subequations}
      \begin{align}
        \pazocal{X} = \{ x \ | \ \exists \ u \text{ s.t. } Wx + Su + p \leq 0 \} \\
        \pazocal{U} = \{ u \ | \ \exists \ x \text{ s.t. } Wx + Su + p \leq 0 \}
      \end{align}
    \end{subequations}
  }
\end{problem}
Note that we do not lose generality without explicit linear equality constraints, since a linear equality constraint can be equivalently formulated with two inequality constraints. The following lemma and proof offers a descriptive solution to this problem.

\begin{lemma}
  \label{lem:lcqpg_sol}
  Given playerwise convexity of objective functions, i.e., $\Gamma_{n}^{uu}$ are positive definite, Problem~\ref{prob:lcqpg} has a piecewise affine feedback Nash equilibrium $u = \phi^\star(x)$ on a finite polyhedral partition of $\pazocal{X}$ if and only if $F$ is invertible.
\end{lemma}

\begin{pf*}{Proof. }
  Assume $S$ has $n_{c}$ rows and we use $E^{n_c}$ to indicate the power set of $\{1, 2, ..., n_{c} \}$.
  For a set of indices $a \in E^{n_{c}}$, we use $W_a$, $S_a$ and $p_a$ to denote picking the corresponding rows. The solution to Problem~\ref{prob:lcqpg} can be found via the following procedure.

  Pick one element $a \in E^{n_{c}}$, which we suppose to be the indices of active constraints, and solve a linear equality constrained quadratic parametric game with the method in Section~\ref{app:lecqpg} as following
  \begin{subequations}
    \begin{align}
      \min_{u_n} \quad & J_n(x, u) = \frac{1}{2}
      \begin{bmatrix}
        1 \\
        x \\
        u
      \end{bmatrix}^\top
      \begin{bmatrix}
        \Gamma_n^{11} & \Gamma_n^{1x} & \Gamma_n^{1u} \\
        \Gamma_n^{x1} & \Gamma_n^{xx} & \Gamma_n^{xu} \\
        \Gamma_n^{u1} & \Gamma_n^{ux} & \Gamma_n^{uu}
      \end{bmatrix}
      \begin{bmatrix}
        1 \\
        x \\
        u
      \end{bmatrix} \\
      \text{s.t. } \quad & W_a x + S_a u + p_a = 0
    \end{align}
  \end{subequations}
  We obtain an affine policy $u = K_a x + s_a$ and the matrices $F$, $P$, $H$ following Lemma~\ref{lem:lecqpg_sol_app}. Note that we are adding subscribes $_a$ to indicate values found associated with the active constraints set of $a$.

  We apply the optimality conditions of VI problem \cite{facchinei2007finite} to find the set in which the policy is an equilibrium in terms of $x$
  \begin{align}  \label{eq:lecqpg_opt_con_app}
    \begin{cases}
      W x + S(K_a x + s_a) + p \leq 0 \\
      -(F (K_a x + s_a) + P x + H) \in \text{cone}\{S^\top_a\}
    \end{cases}
  \end{align}
  where $\text{cone}\{S^\top_a\}$ means the cone generated by the rows of $S_a$. Condition \eqref{eq:lecqpg_opt_con_app} can be reformulated as a polyhedral constraint $L_a x + l_a \leq 0$, since the second condition is equivalent to an linear inequality as shown in Appendix~\ref{app:cone_condition}, where
  \begin{subequations}
    \begin{align}
      L_a &=
      \begin{bmatrix}
        W + S K_a \\
        (S_a S_a^\top)^{-1} S_a (FK_a + P) \\
        [I - S_a^\top (S_a S_a^\top)^{-1} S_a] (FK_a + P) \\
        [- I + S_a^\top (S_a S_a^\top)^{-1} S_a] (FK_a + P)
      \end{bmatrix} \\
      l_a &=
      \begin{bmatrix}
        Ss_a + p \\
        (S_a S_a^\top)^{-1} S_a (Fs_a + H) \\
        [I - S_a^\top (S_a S_a^\top)^{-1} S_a] (Fs_a + H) \\
        [- I + S_a^\top (S_a S_a^\top)^{-1} S_a] (Fs_a + H)
      \end{bmatrix}
    \end{align}
  \end{subequations}
  Hence we have found an affine policy $u = K_a x + s_a$ on a polyhedral region $\pazocal{X}_a \coloneqq \{x \ |\  L_a x + l_a \leq 0\}$ in $\pazocal{X}$.
  In the case of $\pazocal{X}_a = \emptyset$, the policy is not an equilibrium.

  The procedure can be repeated for all combinations of active constraints, i.e., $\forall a \in E^{n_{c}}$. For any $x \in \pazocal{X}$, there exists an equilibrium $u^\star(x)$ and a set of active constraints, therefore $x$ must fall in one of the $\pazocal{X}_a$. Because there are a finite number of combinations of active constraints, a finite number of partitions of $\pazocal{X}$ with $\pazocal{X}_a$ and corresponding equilibria policy in each partition can be found.
  \hfill \qed
\end{pf*}

\subsection{Linearly Constrained Piecewise Quadratic Parametric Game}
We focus on static game problems where each player's cost function $J_n(x, u)$ is continuous, strictly convex and piecewise quadratic on a polyhedral partition $\pazocal{P}$. Due to the complexity of the problem, we refrain from obtaining explicit solutions, but study the properties of the solution in this section.
\begin{problem} \label{prob:lcpqpg}
  {\it
    \textbf{Linearly Constrained Piecewise Quadratic Parametric Game}
    \begin{subequations}
      \begin{align}
        \min_{u_n} \quad & J_{n}(x, u) \\
        \textnormal{s.t. } \quad & z = [x, u] \in \pazocal{P} = \{ \pazocal{Z}_i | i \in \pazocal{I} \} \\
        & \pazocal{Z} = \{[x, y] | W_i x + S_i u + p_i \leq 0 \} \\
        & \pazocal{I} \textnormal{ is an index set}
      \end{align}
    \end{subequations}
  }
\end{problem}
Assume all players share the same polyhedral partition W.L.O.G and that $\pazocal{J}(u)$ is strongly monotone for any given $x$ to guarantee the existence and uniqueness of a solution. In each partition $\pazocal{Z}_i$, the problem reduces to a linearly constrained quadratic parametric game as Problem~\ref{prob:lcqpg}, and $F_i, P_i, H_i$ are inherited from Section~\ref{app:lecqpg}, where the subscript $_i$ denotes the values in $\pazocal{Z}_i$. Suppose there are $n_P$ polyhedral partitions, i.e., $\pazocal{I} = {1, 2, ..., n_P}$. The exact quadratic expression is suppressed.
We define an auxiliary problem for each $i \in \pazocal{I}$.
\begin{problem} \label{prob:lcqpg_aux}
  {\it
    \textbf{Auxiliary problem $\mathbb{P}_i$}
    \begin{subequations}
      \begin{align}
        \min_{u_n} \quad & J_n(x, u) \\
        \textnormal{s.t. } \quad & z \in \pazocal{Z} = \{[x, y] \ |\  W_i x + S_i u + p_i \leq 0  \}
      \end{align}
    \end{subequations}
  }
\end{problem}

Each auxiliary problem can be solved as in Section~\ref{app:lcqpg}, obtaining a piecewise affine policy and piecewise quadratic value functions.
Since Problems~\ref{prob:lcpqpg} and \ref{prob:lcqpg} are closely connected, it is sensible to study the connection between the solutions of these two problems.

We define the concept of \textit{active polyhedrons} $A_{p}([x,u])$ for $[x, u]$ in $\pazocal{P}$ as the set of indices $i \in \pazocal{I}$ such that $\pazocal{Z}_i$ contains $[x, u]$, i.e., $A([x,u]) = \{ i \ | \ W_i x + S_i u + p_i \leq 0 \}$.
The next lemma explains how to recover the solution to Problem~\ref{prob:lcpqpg} from the auxiliary problems.
We use the $\pazocal{U}(x)$ to indicate the set of feasible $u$ with given $x$ and define $\pazocal{U}_i(x) \coloneqq \{ u \ |\ [x, u] \in \pazocal{Z}_i \}$.

The following lemma answers the question of how to solve the equilibrium of Problem~\ref{prob:lcpqpg} for a given $x$ with the help of auxiliary problems Problem~\ref{prob:lcqpg_aux}.

\begin{lemma} \label{lem:lcpqpg_olne}
  $u^\star(x)$ is the open-loop Nash equilibrium to Problem~\ref{prob:lcpqpg} given parameter $x$ if and only if it is the solution to $\mathbb{P}_i, \forall i \in A([x, u^\star(x)])$.
\end{lemma}

\begin{pf*}{Proof. }
  Because for any given $x$ there is a unique solution for Problem~\ref{prob:lcpqpg}, a local solution is also the global solution. The equivalent VI problem for finding the local solution to Problem~\ref{prob:lcpqpg} is
  \begin{subequations}
    \label{eq:local_VI}
    \begin{align}
      \text{find } & u^\star \text{ s.t. } \pazocal{J}(u^\star)^\top(u - u^\star) \geq 0 \\
      & \forall u \text{ in a neighborhood of } u^\star(x) \label{eq:neighborhood}
    \end{align}
  \end{subequations}
  The condition \eqref{eq:neighborhood} can be expanded to $\forall u \in \cup \ \pazocal{U}_i(x), \forall i \in A_p([x, u^\star(x)])$. Due to the same argument that local and global solution are equivalent, this expansion does not change the solution found.
  Therefore, \eqref{eq:local_VI} is equivalent to $u^\star(x)$ being the solution to $\mathbb{P}_i$ simultaneously. We can find the solution of Problem~\ref{prob:lcpqpg} via the solutions of $\mathbb{P}_i$.
  \hfill \qed
\end{pf*}

Next we focus on finding the feedback Nash equilibrium in terms of $x$. Suppose there are a total of $n_{c}$ inequalities. The inequalities defining all partitions can be collected together as
\begin{align}
  W =
  \begin{bmatrix}
    W_1 \\ W_2 \\ \vdots \\ W_{n_{c}} \ \\
  \end{bmatrix} \
  S =
  \begin{bmatrix}
    S_1 \\ S_2 \\ \vdots \\ S_{n_{c}} \  \\
  \end{bmatrix} \
  p =
  \begin{bmatrix}
    p_1 \\ p_2 \\ \vdots \\ p_{n_{c}} \ \\
  \end{bmatrix}
\end{align}
Therefore we can create a one-to-one mapping from an index set $ j \in \{1, 2, ..., n_{c} \}$ to the inequality $W_j x + S_j + p_j \leq 0$.
For any pair of $(x, u)$, we can check the partition(s) it is living in and collect the indices in $A_p([x,u])$, and check the active constraints and collect them in $A_c([x,u])$.
We call $A_c([x,u])$ \textit{acitve inequalities}.
Note that though $A_p([x,u])$ and $A_c([x,u])$ are tied to pairs of $[x,u]$, there could only be a finite number of different $A_p([x,u])$ and $A_c([x,u])$ since there are only a finite number of partitions and inequality constraints.
In other words, we can find a finite number of representative pairs of $[x_k, u_k], k = 1, 2, ..., n_s$, such that $\pazocal{S}_p = \{A_p([x_k, u_k]) \ |\ k \in \{1, 2, ..., n_s\} \}$ and $\pazocal{S}_c = \{A_c([x_k, u_k]) \ |\ k \in \{1, 2, ..., n_s\} \}$ contain all possible active polyhedrons and active constraints combinations.
We can get rid of the dependency on $(x, u)$, use the index $k$ to indicate different feasible active polyhedrons $A_p(k)$ and feasible active constraints $A_c(k)$.
Finding the sets $\pazocal{S}_p$ and $\pazocal{S}_c$ requires studying the structure of the polyhedral partition $\pazocal{P}$, which is not within the scope of this paper.

The next lemma describes the feedback Nash equilibrium of Problem~\ref{prob:lcpqpg}.
\begin{lemma}
  \label{lem:lcpqpg_fne}
  Given playerwise convexity of objective functions, i.e., $\Gamma_{n}^{uu}$ are positive definite, Problem~\ref{prob:lcpqpg} has a piecewise affine feedback Nash equilibrium $u^\star = \phi^\star(x)$ on a finite polyhedral partition of $\pazocal{X}$ if and only if $F_i$ is invertible $\forall i \in \pazocal{I}$.
\end{lemma}

\begin{pf*}{Proof. }
  Similar to the proof for Lemma~\ref{lem:lcqpg_sol}, we describe a procedure for finding affine policies and their corresponding region of $x$ where they are feedback Nash equilibrium. Then we conclude by arguing all $x \in \pazocal{X}$ are included in the procedure.

  We iterate through all $n_s$ pairs of $A_p(k)$ and $A_c(k)$. For each $k$, we can solve for an affine policy $u^\star = K_k x + s_k$ with the feasible active inequalities $A_c(k)$ and the objective functions of all players as they constitute an instance of Problem~\ref{prob:lecqpg}. To find the region where this policy is indeed an FNE, we need to solve the corresponding optimality conditions
  \begin{align}
    \begin{cases}
      W_i x + S_i(K_k x + s_k) + p_i \leq 0, \forall i \in A_p(k) \\
      -(F_i (K_k x + s_k) + P_i x + H_i) \in \text{cone}\{S^\top_{i,a}\}, \forall i \in A_p(k) \label{eq:cone_condition}
    \end{cases}
  \end{align}
  where $S_{i,k}$ means picking the rows of active inequalities in $S_i$ that are also in $A_p(k)$.
  The second condition \eqref{eq:cone_condition} can be expressed with linear inequalities as in Appendix~\ref{app:cone_condition}, therefore the feasible region $\pazocal{X}_k$, if non-empty, is polyhedral, and $u^\star = K_k x + s_k$ is the FNE in this region.
  We omit the explicit expression here.

  This procedure can be repeated $\forall k \in \{1, 2, ..., n_s \}$. For any $x$ in $\pazocal{X}$, we can find $u^\star(x)$ via a solver for monotone VI problems, find the active polyhedrons and active inequalities, therefore the $k$. The solution $[x, u^\star(x)]$ is covered by the condition \eqref{eq:cone_condition} with $k$ because they share the same local optimality condition.
  Note that such procedure may require exponentially many steps w.r.t. the number of polyhedrons and inequality constraints.
  \hfill \qed
\end{pf*}

\subsection{Linearly Constrained Quadratic Dynamic Games}
Now that we are equipped with some basic results, we can move on to dynamic games. A linearly constrained quadratic dynamic game is formulated as following. It is one of the most complicated form of dynamic games of which we can acquire analytical solution in theory. We briefly discuss the dynamic programming solution to such problems.
\begin{problem}
  \label{prob:lq_dynanic_game_app}
  {\it
    \textbf{Linearly Constrained Quadratic Dynamic Games}
    \begin{subequations}
      \begin{align}
        \min_{u_n} \quad & J_n(x, u) = \sum_{k=0}^{T} \frac{1}{2}
        \begin{bmatrix}
          1 \\
          x_k \\
          u_{:,k}
        \end{bmatrix}^\top
        \begin{bmatrix}
          \Gamma_{n,k}^{11} & \Gamma_{n,k}^{1x} & \Gamma_{n,k}^{1u} \\
          \Gamma_{n,k}^{x1} & \Gamma_{n,k}^{xx} & \Gamma_{n,k}^{xu} \\
          \Gamma_{n,k}^{u1} & \Gamma_{n,k}^{ux} & \Gamma_{n,k}^{uu}
        \end{bmatrix}
        \begin{bmatrix}
          1 \\
          x_k \\
          u_{:,k}
        \end{bmatrix} \\
        \text{s.t. } \quad & x_{k+1} = A_k x_k + B_k u_{:,k} + b_k \\
        & W_{n,k} x + S_{n,k} u + p_{n,k} \leq 0
      \end{align}
    \end{subequations}
  }
\end{problem}

The first static problem required to be solved by dynamic programming is a piecewise quadratic, resulting in a piecewise affine policy and piecewise quadratic value functions on $\pazocal{X}_{T}$, which is the space of $x_{T}$.
Based on the linear dynamics and constraints at step $T-1$, we can form a polyhedral partition for $[x_{T-1}, u_{T-1}]$, resulting in the next static game to be a linearly constrained piecewise quadratic parametric game. As we have seen in Section~\ref{app:lcqpg}, the solution remains to be a piecewise affine policy and quadratic value functions.
Therefore, the backward pass by dynamic programming can be done obtaining a series of piecewise quadratic value functions with linear inequality constraints.
However, the number of partitions on each space $\pazocal{X}_k$ can grow exponentially causing the procedure to be computationally prohibiting. It is reasonable to believe that the feedback Nash equilibrium of general constrained dynamic games can be more complex. This fact drives us to seek local FNE.

\end{document}